\documentclass[reprint,
superscriptaddress,
 amsmath,amssymb,
 aps,
showkeys
]{revtex4-1}

\makeatletter
\newcommand\emailx[1]{%
\move@AF%
\def\@affil{{\normalfont\,#1\strut}{}}%
}%

\usepackage{graphicx}
\usepackage{dcolumn}
\usepackage{bm}
\usepackage{hyperref}
\usepackage{physics}
\usepackage{mathtools}
\usepackage{subcaption}

\usepackage{graphicx,amssymb,amsmath,amsthm,amsfonts}
\usepackage[usenames]{color}
\usepackage{mathtools}

\newcommand{\sVert}[1][0]{%
  \ifcase#1\relax
  \rvert\or\bigr|\or\Bigr|\or\biggr|\or\Biggr
  \fi
}

\begin{document}

\preprint{APS/123-QED}

\title{Grand canonical ensemble of a $d$-dimensional \\
Reissner-Nordstr\"om black hole in a cavity}

\author{Tiago V. Fernandes}
\affiliation{Centro de Astrof\'isica e Gravita\c c\~ao - CENTRA,
Departamento de F\'isica, Instituto Superior T\'{e}cnico - IST,
Universidade de Lisboa - UL,\\ Avenida Rovisco Pais 1, 1049-001
Lisboa, Portugal}
\emailx{tiago.vasques.fernandes@tecnico.ulisboa.pt}
\author{Jos\'{e} P. S. Lemos}
\affiliation{Centro de Astrof\'isica e Gravita\c c\~ao - CENTRA,
Departamento de F\'isica, Instituto Superior T\'{e}cnico - IST,
Universidade de Lisboa - UL,\\ Avenida Rovisco Pais 1, 1049-001
Lisboa, Portugal}
\emailx{joselemos@ist.utl.pt}

\date{\today}

\begin{abstract}

The grand canonical ensemble of a $d$-dimensional Reissner-Nordstr\"om
black hole space in a cavity is analyzed in every possible aspect.
The analysis starts with the realization of the grand canonical
ensemble through the Euclidean path integral approach by giving the
Euclidean action for the $d$-dimensional Reissner-Nordstr\"om black
hole with the correct topology, and boundary conditions
corresponding to a cavity. More precisely, the fixed quantities 
of the ensemble are the temperature and the electric potential 
at the cavity boundary. One then performs a zero loop
approximation, to find and analyze the stationary points of the reduced
action. This yields two solutions for the
electrically charged black hole,
the smaller, $r_{+1}$, and the larger, $r_{+2}$. 
Through perturbations of the reduced action 
around the stationary points, one finds stability criteria for the 
solutions of the black hole that show that $r_{+1}$ is unstable and 
$r_{+2}$ is stable. Moreover,
one analyzes the most probable
configurations for each value of the fixed quantities at the boundary, 
with the configurations being either a stable charged black hole or 
hot flat space. One also compares the stable black hole with 
a nongravitating charged shell, which serves as a model for 
an electrically
charged hot flat space. By making
the correspondence between the action already evaluated and the grand
canonical ensemble potential of thermodynamics one can get the
entropy, the mean charge, the mean energy, and the thermodynamic
pressure, as well as the
Smarr formula, here shown to be
valid only for the unstable black hole
$r_{+1}$. We make a stability analysis in terms of thermodynamic
variables, which yields that thermodynamic stability is related 
to the positivity of the heat capacity at constant electric potential 
and constant area of the cavity. We also  comment on
the most favorable thermodynamic phases
and deduce the possible phase
transitions. We then pick up a specific dimension, $d=5$,
which is singled
out naturally from the other higher dimensions
as it provides an exact solution for the problem,
and apply
all the results previously found.
The case $d=4$ is concisely put in an appendix where the
results are directly equated with previous
works.  We also compare thermodynamic radii with the photonic
orbit radius and the Buchdahl-Andr\'easson-Wright bound radius in
$d$-dimensional Reissner-Nordstr\"om spacetimes and find
they are unconnected, showing that the connections
displayed in the Schwarzschild case are not generic,
rather they are very restricted equalities
holding only in the pure gravitational situation.

\end{abstract}

\maketitle


\section{Introduction}
\label{sec:Intro}

Black hole thermodynamics has been studied since the hypothesis of
Bekenstein that a black hole has entropy \cite{Bekenstein:1973}, and
since the four laws of black hole mechanics have been introduced
\cite{Bardeen:1973}.  It was put on a firm ground after Hawking
discovered that black holes radiate field quanta with a thermal
spectrum at temperature $T_{\rm H} = \frac{\kappa}{2\pi}$, the Hawking
temperature in Planck units, where $\kappa$ is the usual black hole
surface gravity \cite{Hawking:1975}.  Moreover, through path integral
methods, it was shown that the correct vacuum state that sits at the
black hole horizon and enables the radiation to be produced is the
Hartle-Hawking vacuum state \cite{Hartle:1976}.  Taking all these
results together, it was possible to deduce that a black hole is indeed
a thermodynamic object with entropy
$S=\frac{A_+}{4}$, the Bekenstein-Hawking entropy,
where $A_+$ is the black hole  horizon area.

A proper construction of the thermodynamics of a black hole can be
done by building a statistical ensemble for the space under analysis.
In the canonical and grand canonical ensembles one has to find the
partition function $Z$ of the corresponding space.  A powerful
approach to find $Z$ is the Euclidean path integral approach to
quantum gravity.  With this method one finds that the partition
function of the space under analysis is given by the path integral
over the possible Euclidean metrics $g$ and fields $\phi$ of the
Euclidean action $I[g,\phi]$, with the restriction that the space and
the fields are periodic with imaginary time length at the boundary
being the inverse temperature, i.e., $Z = \int Dg\,D\phi\,\,{\rm
e}^{-I[g,\phi]}$.  To make progress in the computation of the
partition function, one can use a zero loop approximation, where the
only contribution to the path integral is the configuration which
minimizes the action $I$ with respect to the relevant parameters.
This action is the classical action and the partition function is then
$Z= {\rm e}^{-I}$.  In spite of $Z$ being constituted by the classical
action alone in this approximation, it vanishes in the classical
limit, i.e., it disappears when the Planck constant goes to
zero. Thus, the partition function $Z$ still has a quantum
gravitational character, indeed one is dealing with the semiclassical
approximation.  The partition function of the canonical or grand
canonical ensembles can then be related to the thermodynamic canonical
or grand canonical potential and the thermodynamics of the space and
fields can be worked out.
These ideas were applied with some success to the Schwarzschild and
Reissner-Nordstr\"om black holes in which the canonical and grand
canonical ensembles, respectively, were formulated with heat reservoir
boundaries with infinite radius, i.e., at infinity \cite{GH77}.
There, it was shown that the configuration which corresponds to a
stationary point of $I$ consisted in a black hole with
Hawking temperature and Bekenstein-Hawking entropy.  However, it was
also shown that, at least for the Schwarzschild black hole, the zero
loop approximation is not valid, since this configuration has negative
heat capacity, indeed it is an extremum that does not minimize the
action $I$.  In fact, in this set up the Schwarzschild solution
corresponds to a saddle point of $I$ and thus acts as an instanton to
hot flat space \cite{Gross:1982}.  Remarkably, it was later found that
if the boundary, at which a temperature is defined, was put at a finite
radius, specifically, at a radius less or equal to the circular photon
orbit radius, then the instability would disappear \cite{allen}.

York \cite{York:1986} then realized that a black hole space in a
spherical cavity at finite radius attached to a heat reservoir at
finite temperature was the correct setup to construct the canonical or
grand canonical ensembles, and through the path integral approach it
was possible to obtain sensible results. In this scheme, for the
Schwarzschild black hole there are two stationary points of the action
$I$. The one which has the least mass is unstable and corresponds to
the Gibbons-Hawking black hole when the radius of the cavity tends to
infinity. The other, which has the highest mass, is stable.  Further
developments to understand the Schwarzschild black hole geometry in
the quantum gravity context of York's formalism were performed in
\cite{whitingyork}, the inclusion of matter fields within the
formalism was sketched in \cite{Martinez:1989}, and the relation
between the various statistical physics ensembles was studied in
\cite{bcmmwhitingyork}.  Using York's formalism, the grand canonical
ensemble of a Reissner-Nordstr\"om black hole in a cavity, where the
temperature and the electric potential are specified at the boundary,
was obtained in \cite{Braden:1990}.  An important development was
achieved in \cite{Zaslavskii:1990} where the canonical ensemble for
arbitrary configurations of self-gravitating systems was analyzed.
In general relativistic asymptotically
anti-de Sitter spacetimes with black holes,
the canonical ensemble was
studied in three and four dimensions \cite{browcreima},
in the two-dimensional Teitelboim-Jackiw theory
the corresponding 
asymptotically
anti-de Sitter  black hole
was analyzed in the canonical ensemble
\cite{lemos1996},
and the  grand canonical ensemble
of the anti-de Sitter Reissner-Nordstr\"om black
hole in four dimensions in general relativity
was put forward in \cite{Peca:1999}. 
Several studies in higher dimensions have also been
performed. Indeed, 
the stability and the negative mode for a higher-dimensional
Schwarzschild black hole in a finite cavity was studied in
\cite{gregoryross}, the canonical ensemble of Schwarzschild-anti-de
Sitter black holes was done in \cite{Akbar:2010}, the $d=5$ and
generic $d$ Schwarzschild black holes were analyzed in
\cite{Andre:2020,Andre:2021}.  Gravastars in the canonical ensemble
were studied in \cite{miyashita}.
The Euclidean path integral approach had its domain
of study extended from
black holes to black branes.  For instance, it has been conjectured
that the mechanical stability of black branes is interwoven with their
local thermodynamic stability \cite{Gubser:2001}.  This conjecture has
been analyzed and proved in certain cases, see
\cite{Reall:2001,Miyamoto:2007,collingbourne}.

It is worth noting that there are similarities in York's formalism and
the analysis of the thermodynamics of a hot thin shell of matter.
This has been established for thin shells with an outer Schwarzschild
spacetime in \cite{Martinez:1996} and for thin shells with an outer
Reissner-Nordstr\"om spacetime
\cite{Lemos:2015a}.  Hot thin shells were then studied in
higher dimensions in a Schwarzschild-Tangherlini spacetime in
\cite{Andre:2019} and in a Reissner-Nordstr\"om-Tangherlini
spacetime in \cite{Fernandes:2022}.
The Reissner-Nordstr\"om case has been revisited in
\cite{bergliaffa2}.

In this work, we generalize the work in \cite{Braden:1990} for an
arbitrary number of dimensions, and we generalize the work in
\cite{Andre:2020,Andre:2021} by including electric charge.  Thus, we
construct the grand canonical ensemble of a Reissner-Nordstr\"om space
in a cavity for an arbitrary number of dimensions $d$.  We compute the
two possible solutions and analyze their stability, by looking at the
reduced Euclidean action. We obtain the thermodynamic properties of
the stable solution and comment about the similarities between this
treatment and the analysis of the thermodynamics of a charged
spherical thin shell in an arbitrary number of dimensions.
We also compare the important thermodynamic radii
that arise in the grand canonical ensemble with
the
photonic orbit radius and the Buchdahl-Andr\'easson-Wright 
bound radius \cite{Wright:2015} of
$d$-dimensional Reissner-Nordstr\"om spacetimes
to confirm or dispel some coincidences. In the
thermodynamic investigation
of stability we use the analysis set in \cite{reif}.

This paper is organized as follows.
In Sec.~\ref{sec:PathIntApp}, we describe the path integral approach
and apply it to the $d$-dimensional Reissner-Nordstr\"om
black hole space in a
cavity, stipulating the statistical mechanics grand
canonical partition function $Z$ which is
found from the path integral of the action  
$I$.
In Sec.~\ref{zerolappp}, we perform the zero loop approximation and
obtain the reduced action, the stationary points of the reduced action,
make a stability analysis, and look for the most probable
configurations.
In Sec.~\ref{sec:GrandCanonical}, we study the system as a
grand canonical thermodynamic system, provide the connection between the
grand canonical potential and the statistical mechanics grand
canonical partition function $Z$ found from the action
$I$ of the path integral,
analyze the thermodynamic quantities and relations, perform a
stability analysis in terms of thermodynamic variables, and look for
the most favorable thermodynamic phase and phase transitions.
In Sec.~\ref{d=5}, we apply the analysis to the case of $d=5$, 
which possesses an explicit exact formula for the stationary points, 
and we study in detail the
zero loop approximation and the thermodynamics with plots.
In Sec.~\ref{sec:conc}, we present the conclusions and further discussions. 
There are several appendices which supplement the analysis of the paper.
In the Appendix~\ref{photonandbhcdahlandreasson}, we study the
photonic orbit radius and the Buchdahl-Andr\'easson-Wright 
bound in
$d$-dimensional Reissner-Nordstr\"om spacetime to compare with the
grand canonical ensemble featured radii.
In the Appendix~\ref{app:gradactioncavity}, we study the smoothness of
the critical points, namely, hot flat space and extremal black hole.
In the Appendix~\ref{canthermo}, we make further comparisons between
the grand canonical ensemble of statistical mechanics and the
grand potential of thermodynamics.
In the Appendix~\ref{app:d4case}, we apply to the case of $d=4$ and
compare  with the results of previous works.
In the Appendix~\ref{photonsphere}, we derive the null geodesic sphere,
i.e., the circular photon orbit radius, in the Reissner-Nordstr\"om
geometry, giving its value in terms of the horizon radius, the
electric potential there, and the spacetime dimension, so that 
we can compare it to the thermodynamic radii. We assume natural units, i.e.,
$G=1$, $c=1$, and $\hbar=1$.

\section{Grand canonical ensemble through the path
integral approach}
\label{sec:PathIntApp}

\subsection{The path integral and thermodynamics}

Assume a
system  comprised of a $d$-dimensional spacetime $\mathcal{M}$ with
associated metric $\bm{g}$ and Maxwell vector field $\bm{A}$. The spacetime
$\mathcal{M}$ can be foliated by the history of a spacelike
hypersurface $\Sigma_t$ with induced metric $\bm{h}$, where $t$ is the
parameter of a congruence of timelike curves that intersect
$\Sigma_t$. We can consider also that each $\Sigma_t$ has a boundary
$\partial \Sigma_t$, and we can build the history of $\partial
\Sigma_t$, which we denominate $\mathcal{B}$, with an induced metric
$\bm{\gamma}$.  The path integral approach states that the evolution
of the system from one state to another is given by the path integral
$\int D[\bm{g}]D[\bm{A}] {\rm e}^{i I_L[\bm{g},\bm{A}]}$, where
$I_L[\bm{g},\bm{A}]$ is the Lorentzian action.

The Euclidean path integral approach of gravitational statistical
mechanics and thermodynamics, which we follow, prescribes, first, an
analytic extension of the spacetime to a Euclidean space with a Wick
transformation of the time, namely, $t=-i\tau$.  Second, it states
that the partition function of this Euclidean space is given by
\begin{align}
Z(\beta,\phi,\bm{\gamma}) = \int D[\bm{g}]D[\bm{A}] {\rm e}^{-
I[\beta,\phi,\bm{\gamma}; \bm{g},
\bm{A}]}\,,\label{eq:Zpathintegral}
\end{align}
where $I[\beta,\phi,\bm{\gamma}; \bm{g}, \bm{A}]$ is the Euclidean
action of the system, $D[\bm{g}]$ and $D[\bm{A}]$ are the respective
measures of the integral for the Euclidean metric and the Maxwell
field, $\beta$ is the inverse temperature at $\mathcal{B}$ and $\phi$
is the electrostatic potential at $\mathcal{B}$, both of which will
be defined below.

When performing the integral, there are several considerations one
must take into account,
as the notation in Eq.~\eqref{eq:Zpathintegral} is too
abstract. First, the path integral has the requirement that the
Euclidean metric is periodic in time. This is motivated by the usual
construction of the partition function of large quantum systems. The
path integral in $\bm{g}$ is the sum of all the paths, i.e.,
of all the
possible values of $\bm{g}$, starting at a hypersurface
$\Sigma_{\tau_{0}}$ and then returning to $\Sigma_{\tau_0}$, where for
simplicity we adopt $\tau_0 = 0$. Note that the path integral is also
summing all the possible induced metrics in $\Sigma_{t_0}$. Second,
the path integral needs to be performed taking into consideration
fixed data at
the boundary $\mathcal{B}$. The fixed data we are considering is the
inverse temperature $\beta$ given by
\begin{align}
\beta = \int_0^{2\pi} (\gamma_{\tau \tau})^{1/2} d\tau\,,
\end{align}
which is also the proper Euclidean
time length of $\mathcal{B}$, by construction
of the partition function, with $\gamma_{\tau \tau}$
being the time-time component of the induced
metric $\bm{\gamma}$.
In addition, the components of the vector field
$A^{a}$ at $\mathcal{B}$ need to be fixed. More specifically, we
define the electrostatic potential at $\mathcal{B}$ as
\begin{align}
\beta \phi = 2\pi i A_{\tau}\sVert[1]_{\mathcal{B}}.
\label{maxinit}
\end{align}
In the case the integral is well-defined and can be performed,
the partition
function $Z$ of Eq.~\eqref{eq:Zpathintegral}
will just depend on the inverse temperature $\beta$, the
electrostatic potential $\phi$ and other components of the induced
metric $\bm{\gamma}$.

We can relate the partition function of the grand canonical ensemble
directly to the thermodynamic grand potential and therefore obtain the
thermodynamics of the system. Note
also that one can instead obtain the
thermodynamic quantities via the usual derivatives of $\ln Z$. The
partition function can be written as
$
Z(\beta,\phi,\bm{\gamma})_{\rm GC} = {\rm e}^{-\beta W[\beta, \phi,
\bm{\gamma}]}$,
where $W$ is the grand potential.

\subsection{The Euclidean action}

The Euclidean action of the system considered here is the Euclidean
Hilbert-Einstein-Maxwell action given by
\begin{align}
   I =& - \int_{\mathcal{M}}\left(\frac{R}{16\pi} -
   \frac{F_{ab}F^{ab}}{4}\right)\sqrt{g} d^dx \nonumber\\&-
   \frac{1}{8\pi}\int_{\mathcal{B}} (K-K_0) \sqrt{\gamma}
   d^{d-1}x\,,
   \label{eq:EucAction1}
\end{align}
where $R$ is the Ricci tensor depending on the metric $g_{ab}$ and its
first and second derivatives,
$g$ is the determinant of $g_{ab}$,
$F_{ab}= \nabla_a A_b - \nabla_b
A_a$ is the strength field tensor for the Maxwell field $A_a$,
$\nabla_a$ is the covariant derivative compatible with $g_{ab}$, $K$ is
the trace of the extrinsic curvature of $\mathcal{B}$ and $K_0$ is the
trace of the extrinsic curvature of $\mathcal{B}$ embedded in flat
space. We use the convention that given an outward unit normal
vector $n^a$ to the hypersurface $\mathcal{B}$, then $K = \nabla_a
n^a$.

\subsection{Topology and boundary conditions}

We are interested in evaluating the path integral with the boundary
$\mathcal{B}$ described by the history of a static spherical
surface. It
is convenient to assume that the dominant terms in the path integral
will correspond to sets of metrics that are spherically symmetric.
For coordinates $(\tau,y,\theta^A)$, where $\theta^A$ are 
the $d-2$ spherical coordinates of the space, 
we assume a line element of the form 
\begin{align}
ds^2 = b^2(y) d\tau^2 + \alpha^2(y)dy^2 +
r^2(y)d\Omega^2\,,
\label{eq:metric}
\end{align}
where $b(y)$, $\alpha(y)$, and $r(y)$, are the metric functions,
functions 
of $y$ alone, and
$d\Omega$ is the volume element of a $(d-2)$-sphere. The
volume element of the space is $\sqrt{g}d^dx= b \alpha r^{d-2}
d\tau dy d\Omega$.

We impose boundary conditions to the metric. At $y=0$, we require
\begin{align}
&b(0)=0\,,\label{eq:boundb}\\
& \eval{(b'\alpha^{-1})}_{y=0} = 1\,,\label{eq:boundbprime}
\end{align}
where $b' = \frac{d b}{d y}$. The first condition comes
from $y=0$ being a degenerate hypersurface, since it corresponds to
the black hole horizon. The degenerate hypersurface will have a
geometry $\mathbb{S}_{d-2}$, i.e.,
it is the geometry of a $d-2$-sphere. Notice
that hypersurfaces of constant $y$ have a geometry $\mathbb{S}_1\times
\mathbb{S}_{d-2}$, in which the length of $\mathbb{S}_1$ smoothly goes
to zero as $y$ goes to zero. The second condition is required so that
the geometry is regular, i.e., there is
no conical
singularity at $y=0$.
Defining $r' = \frac{d r}{d y}$, 
we also require that
\begin{align}
\eval{\left(\frac{r'}{\alpha}\right)}_{y=0} = 0\,,
\label{eq:EHtopology}
\end{align}
so that the hypersurface at $y=0$ inherits the property that the
normal to the black hole horizon $\eval{n_a=(0,r',0,0)}_{y=0}$ is a
null vector. In even dimensions, the condition given in
Eq.~\eqref{eq:EHtopology} can be motivated by requiring that
$\mathcal{M}$ has an Euler number $\chi = 2$, but in odd dimensions,
there is no equivalent motivation, and since there is no
effective difference between even and odd dimensions
in this context, it is better to generically
stick to the inheritance of the
normal to the black hole horizon property. Finally, we  define that
$r(0)=r_+$ is the radius of the black hole horizon. The
parameter $r_+$ is not fixed in the path integral.

The boundary $\mathcal{B}$ is
assumed to be located at $y=1$ with induced metric
\begin{align}
ds_\mathcal{B}^2 = b(1)^2 d\tau^2 + R^2 d\Omega^2\,, 
\end{align}
and its volume element
is $\sqrt{\gamma}d^{d-1}x = b(1)R^{d-2}d\tau
d\Omega$, where we have defined $R\equiv r(1) $.
The unit normal vector to the
boundary $\mathcal{B}$ is $r_a = (0,\alpha,0,0)$.
The $\tau={\rm constant}$ surfaces have area $A$ given by
\begin{align}
A=\Omega R^{d-2}\,,
\label{areaA}
\end{align}
where $\Omega$ is the area of the unit sphere
in $d-2$ dimensions $\Omega=\frac{2\pi^{\frac{d-1}{2}}}
{\Gamma\left(\frac{d-1}{2}\right)}$.
For $d=4$, the area is $\Omega=4\pi$, for $d=5$,
$\Omega=2\pi^2$, and so on for other higher dimensions.

The condition of
fixed temperature, or fixed
inverse temperature, at the boundary $\mathcal{B}$ is given by the
boundary condition
\begin{align}
\beta = 2\pi b(1)\,,\label{eq:betadef}
\end{align}
where $\beta$ is the inverse temperature, i.e.,
$\beta=\frac1T$, with $T$ being the temperature  at the boundary. 
For the Maxwell field $A_a$, spherical
symmetry requires that the only nonzero component of the strength
field tensor is $F_{\tau y}$. One can choose a gauge in which only the
component $A_\tau$ of the Maxwell field is nonzero. It must be
of the form
$A_\tau = A_\tau(y)$. We require that the Maxwell field is bounded,
and so we impose that
\begin{align}
A_{\tau}(0) = 0\,.\label{eq:Ay0}
\end{align}
At the boundary $\mathcal{B}$, the Maxwell field should obey the
boundary condition motivated above
\begin{align}
\beta \phi = 2\pi i A_{\tau}(1)\,,
\label{eq:phidef}
\end{align}
see Eq.~\eqref{maxinit}.

\subsection{Action of the spherically symmetric space}

The action given in Eq.~\eqref{eq:EucAction1} of
the space with the given boundary conditions can now be 
computed.

The first integrand in the action corresponding to the
integral over $\mathcal{M}$ is the Ricci scalar $R$. For
the line element given in Eq.~\eqref{eq:metric},
it can
be written as
\begin{align} -\sqrt{g}\frac{R}{16\pi} =
\frac{1}{8\pi}\left(\frac{r^{d-2}b'}{\alpha}\right)' +
\frac{\alpha b
r^{d-2}}{8\pi} {G^{\tau}_{}}_{\tau}\,,\label{eq:Riccimetric}
\end{align}
where ${G^{\tau}_{}}_{\tau}$ is the time-time
component of the Einstein tensor
${G}_{ab}$, given by
\begin{align} {G^{\tau}_{}}_{\tau} = \frac{d-2}{2 r'
r^{d-2}}\left(r^{d-3}\left[\left(\frac{r'}{\alpha}\right)^2 -
1\right]\right)'\,,
\label{eq:Gtt}
\end{align}
and a prime means derivative with respect to $y$.
The second integrand is given by
\begin{align}
\sqrt{g}\frac{F_{ab}F^{ab}}{4} = \frac{1}{2} \frac{r^{d-2}}{\alpha
b} (A'_{\tau})^2\,,
\end{align}
where the expression $F_{y\tau} = A'_\tau$ was used. 
The third integrand is related to the
extrinsic curvature of a hypersurface with constant $y$.
The
extrinsic curvature of such a hypersurface is 
$\bm{K} = \frac{b b'}{\alpha} d\tau^2 
+ \frac{r r'}{\alpha} d\Omega^2$, 
while the extrinsic curvature of a hypersurface with constant $y$
embedded in flat space 
is just $\bm{K}_0 = r d\Omega^2$, 
since in flat space $\frac{r'}{\alpha_0} = 1$ and 
$b_0$ is constant. Therefore, the integrand of the integral 
over the boundary $\mathcal{B}$ is
\begin{align}
&-\frac{\sqrt{\gamma}}{8\pi}(K-K_0) =\nonumber\\&
\eval{\frac{1}{8\pi}\left[ (d-2)br^{d-3}\left(
1- \frac{r'}{\alpha}\right) -
\frac{r^{d-2}b'}{\alpha}\right]}_{y=1}\,.
\label{boundterm}
\end{align}

Having explicitly showed the integrands
of the action given in Eq.~\eqref{eq:EucAction1},
we can perform their integration.
Let us present some simplifications in
the integration process. 
The integral 
of the first term in Eq.~\eqref{eq:Riccimetric} becomes two 
terms which are $\eval{\frac{r^{d-2}b'}{8\pi\alpha}}_{y=1} 
- \eval{\frac{r^{d-2}b'}{8\pi\alpha}}_{y=0}$, i.e., a boundary 
term at $y=1$ and another at $y=0$. Moreover, the boundary term 
at $y=1$ cancels with the term 
$-\eval{\frac{r^{d-2}b'}{8\pi\alpha}}_{y=1}$ in Eq.~\eqref{boundterm}.
To evaluate
the term 
$\eval{\frac{r^{d-2}b'}{8\pi\alpha}}_{y=0}$
we use the boundary condition Eq.~\eqref{eq:boundbprime}, 
which is the condition
that requires the absence of
conical singularities, to 
rewrite the boundary term at $y=0$ as 
$\eval{\frac{r^{d-2}b'}{\alpha}}_{y=0} = r_+^{d-2}$.
Moreover, the term 
$\eval{(d-2)br^{d-3}\left(
1- \frac{r'}{\alpha}\right)}_{y=1}$
in Eq.~\eqref{boundterm}
can be further simplified.
Integrating it over the Euclidean time
coordinate which has period $2\pi$ and 
 using 
the boundary condition in Eq.~\eqref{eq:betadef}, i.e.,
$\beta = 2\pi b(1)$, one obtains 
$(d-2)
\beta R^{d-3}\left(1- \eval{\frac{r'}{\alpha}}_{y=1}\right)$.

Performing then the final integrations, one finds the action
\begin{align}
&I[\beta,\phi,R;b,\alpha,r,A_\tau] = \frac{(d-2)\Omega}{8\pi}
\beta R^{d-3}\left(1- \eval{\frac{r'}{\alpha}}_{y=1}\right)\nonumber \\
& - \frac14 \Omega r_+^{d-2} + \int_{\mathcal{M}}\frac{\alpha b
r^{d-2}}{8\pi}{G^{\tau}_{}}_{\tau}d^dx 
+ \int_{\mathcal{M}}\frac{r^{d-2}}{2\alpha b}(A'_\tau)^2
d^dx\,.
\label{eq:rawaction}
\end{align}
Formally, one would then proceed with the calculation of the path
integral in $b$, $\alpha$, $r$ and $A_\tau$.

\section{Zero loop approximation}
\label{zerolappp}

\subsection{The reduced action}

We perform the zero loop approximation of the path integral with the
action in Eq.~\eqref{eq:rawaction}, since we are interested in the
semiclassical computation of the partition function. The zero loop
approximation consists on considering only the contribution of the
classical path to the path integral, neglecting the other paths. The
classical path obeys in this case the Einstein-Maxwell equations with
the prescribed boundary conditions. Note that there may be more than
one classical path, as we will see.
Here, we follow the procedure given in \cite{Braden:1990}. First, one
imposes the constraint equations of the Einstein-Maxwell equations to
the possible paths. Then, one finds the zero-order action
and the semiclassical partition function.

The constraints of the metric are the momentum constraint
and the Hamiltonian constraint that come out
from the Einstein equation $G_{ab}=8\pi T_{ab}$, where
$T_{ab}$ is the electromagnetic field energy-momentum tensor.
The momentum constraint
is satisfied since the extrinsic curvature of a $\tau={\rm const}$
hypersurface vanishes. The Hamiltonian constraint is given by
${G^\tau}_{\tau} = 8\pi {T^{\tau}}_\tau$, where
$T_{ab} = F_{ac}F_{bd}g^{cd} - \frac{1}{4}g_{ab}F_{cd}F^{cd}$.
So,
the Hamiltonian constraint is
\begin{align}
{G^\tau}_{\tau}=
\frac{4\pi}{r^{2(d-2)}}
\left(\frac{r^{d-2}}{\alpha b} F_{y\tau}\right)^2
\,.
\label{eq:HamiltonianConst}
\end{align}
The Gauss constraint 
for the Maxwell field is given by the equation
\begin{align}
\nabla_y{{{F^{\tau y}}_{}}_{}} = 0\,.
\label{gaussc}
\end{align}
One can integrate the Gauss constraint, Eq.~\eqref{gaussc},
to find $\frac{r^{d-2}}{\alpha b} F_{y\tau}  = -
i \frac{q}{\Omega}$,
where $q$ is
a constant of integration
chosen to be the Lorentzian electric charge of the system.
Then, using Eq.~\eqref{eq:Gtt} for ${G^\tau}_{\tau}$
in Eq.~\eqref{eq:HamiltonianConst},
the Hamiltonian constraint is 
$\frac{(d-2)}{2 r'
r^{d-2}}\left(r^{d-3}\left(\frac{r'^2}{\alpha^2}-1\right) \right)'
= -\frac{4\pi}{\Omega^2
r^{2d-4}}q^2$.
The Hamiltonian constraint can be integrated to 
 yield
\begin{align}
\left( \frac{r'}{\alpha}\right)^2 = 1 - \frac{2\mu m}{r^{d-3}} +
\frac{\lambda q^2}{r^{2d-6}}\,,
\label{eq:alphavalue}
\end{align}
where $\mu m$ is a constant
of integration and $\lambda$ is a convenient parameter.
The constant 
$\mu m$ can be written
with the help of the boundary condition in
Eq.~\eqref{eq:EHtopology} as
\begin{align}
2\mu m = r_+^{d-3} + \frac{\lambda q^2}{r_+^{d-3}}\,.
\label{eq:mumr+q}
\end{align}
When $m$ is understood as a mass then
\begin{align}
\mu = \frac{8\pi}{(d-2)\Omega}\,,
\label{eq:mu}
\end{align}
such that in $d=4$ one has  $\mu=1$, 
and the mass term is $2m$.
The parameter $\lambda$ is given by
\begin{align}
\lambda = \frac{8\pi}{(d-2)(d-3)\Omega^2}\,,
\label{eq:lambda}
\end{align}
such that in $d=4$, $\lambda=\frac1{4\pi}$.
Since $F_{y\tau} = A'_\tau$, the integral of the Gauss constraint,
Eq.~\eqref{gaussc}, is
\begin{align}
\frac{r^{d-2}}{\alpha b} A'_{\tau} = -
i \frac{q}{\Omega}\,.
\label{eq:MaxwellA}
\end{align}

The action
given in
Eq.~\eqref{eq:rawaction} can now be simplified using
Eqs.~\eqref{eq:alphavalue} and
\eqref{eq:MaxwellA} in the following
way.
For the first term in Eq.~\eqref{eq:rawaction}, we use
Eq.~\eqref{eq:alphavalue} to obtain
\begin{align}
\left(1- \eval{\frac{r'}{\alpha}}_{y=1}\right)
=1-\sqrt{f[R;r_+,q]} \,,
\label{fd0}
\end{align}
where $f[R;r_+,q]$ is defined as
\begin{align}
&f[R;r_+,q] =
\left(1-\frac{r_+^{d-3}}{R^{d-3}}\right)\left(1-\frac{\lambda
q^2}{r_+^{d-3} R^{d-3}}\right)\,,
\label{fd}
\end{align}
and where use of Eq.~\eqref{eq:mumr+q} has been made.
The term
$\int_{\mathcal{M}}\frac{\alpha b
r^{d-2}}{8\pi}{G^{\tau}_{}}_{\tau}d^dx$ 
in Eq.~\eqref{eq:rawaction} can also be
simplified using Eq.~\eqref{eq:HamiltonianConst},
$F_{y\tau} = A'_\tau$, 
and Eq.~\eqref{eq:MaxwellA}, yielding
\begin{align}
\int_{\mathcal{M}}\frac{\alpha b
r^{d-2}}{8\pi}{G^{\tau}_{}}_{\tau}d^dx &= \int d\tau dr d\Omega
\left(\frac{r^{d-2}}{2\alpha b}A'_{\tau}\right)A'_{\tau}
\nonumber\\&= - \frac{1}{2}q\beta \phi\,.
\end{align}
The term depending explicitly on $A'_\tau$
in Eq.~\eqref{eq:rawaction}, namely,
$\int_{\mathcal{M}}\frac{r^{d-2}}{2\alpha b}(A'_\tau)^2
d^dx$, 
can also be simplified.
Using
Eqs.~\eqref{eq:Ay0}, \eqref{eq:phidef},
and \eqref{eq:MaxwellA}, one finds
\begin{align}
\int_M \left(\frac{r^{d-2}}{2\alpha b}A'_\tau\right)A'_\tau d^dx
&= \int d\tau d\Omega
\left(-i\frac{q}{2\Omega}\right)\eval{A_\tau}_{y=1} \nonumber\\&=
- \frac{1}{2}q\beta\phi \,.
\label{lastint}
\end{align}
Then, putting Eqs.~\eqref{fd0}-\eqref{lastint} into
the action of 
Eq.~\eqref{eq:rawaction}, we obtain the reduced
action $I_*$ as
\begin{align}
I_*[\beta,\phi, R; r_+,q] =& \frac{(d-2)\Omega
R^{d-3}\beta}{8\pi}\left(1-\sqrt{f[R;r_+,q]}\right) \nonumber\\&-
q \beta \phi - \frac{\Omega r_+^{d-2}}{4}\,,
\label{eq:reducedaction}
\end{align}
Note that the parameters $r_+$ and $q$ are not fixed in the path
integral. In fact, with the constraints applied and the spherical
symmetry of the system, the remaining paths correspond to spaces with
$\tau={\rm const}$ slices parametrized by any possible value of $r_+$
and $q$. The partition function with the constraints is then given by
\begin{align}
Z[\beta,\phi,R] = \int D[r_+]D[q] {\rm e}^{-I_*[\beta,\phi,R;r_+,q]}\,.
\label{ZasI}
\end{align}
To fully apply the zero loop approximation, we must impose the rest of
the Einstein-Maxwell equations. This turns out to be equivalent to finding
the stationary points of the reduced action in $r_+$ and $q$. One then
must analyze if the stationary points minimize the action, since only
there the zero loop approximation is valid. The motivation for
applying first the constraint equations instead of applying the full
equations is that the reduced action is useful to understand the
validity of the zero loop approximation and therefore the stability of
the solution given by the stationary point.

\subsection{Stationary points of the reduced action}
\label{sec:extremastability}

\subsubsection{Stationary points of the reduced action
properly said}

The solutions for an electrically charged
black hole in a cavity 
within a reservoir are given
by finding the
stationary points of the reduced action
given in Eq.~\eqref{eq:reducedaction}, specifically,
\begin{align}
&\frac{\partial I_*}{\partial r_+} = 0\,,\label{eq:derI1}\\
&\frac{\partial I_*}{\partial q} = 0\,.\label{eq:derI2}
\end{align}
Using  Eq.~\eqref{eq:reducedaction}
together with
Eqs.~\eqref{eq:derI1} and\eqref{eq:derI2},
one finds that the
stationary points of $I_*[\beta,\phi, R; r_+,q]$
occur when $\beta$ and $\phi$
assume the following expressions
\begin{align}
&\beta = \frac{4\pi}{(d-3)}\frac{r_+^{2d-5}}{r_+^{2d-6}-\lambda
q^2}\sqrt{f[R,r_+,q]}\,,\label{eq:betaexpress}\\
&\phi = \frac{q}{(d-3)\Omega
\sqrt{f[R,r_+,q]}}\left(\frac{1}{r_+^{d-3}}-
\frac{1}{R^{d-3}}\right)\,,
\label{eq:phiexpress}
\end{align}
respectively.
Since
in the canonical ensemble
$\beta$ and $\phi$ are fixed, the solutions for the system of
equations are $r_+= r_+(\beta,\phi,R)$ and $q=q(\beta,\phi,R)$.
The reduced
action evaluated at a stationary point in $r_+$ and $q$
is then formally
\begin{align}
I_0[\beta,\phi,R]=
I_*[\beta,\phi,R;r_+[\beta,\phi,R],q[\beta,\phi,R]]
\,.
\label{eq:actionI0}
\end{align}
From Eq.~\eqref{eq:reducedaction} one can thus write
the reduced action $I_0[\beta,\phi,R]$ of Eq.~\eqref{eq:actionI0}
as 
\begin{align}
&I_0[\beta,\phi,R] =\nonumber\\
& \frac{(d-2)\Omega R^{d-3}\beta}{8\pi}\left(1- 
\sqrt{f[R;r_+[\beta,\phi,R],q[\beta,\phi,R]]}\right)
\nonumber
\\
& -q[\beta,\phi,R] \beta \phi -
\frac{\Omega r_+^{d-2}[\beta,\phi,R]}{4}\,,
\label{eq:actionI0full}
\end{align}
where $f[R;r_+[\beta,\phi,R],q[\beta,\phi,R]]$
can be taken from Eq.~\eqref{fd} and the
solutions of Eqs.~\eqref{eq:betaexpress}
and \eqref{eq:phiexpress}.
The
partition function of the system given in Eq.~\eqref{ZasI}
then reduces to
\begin{align}
Z[\beta, \phi, R] = {\rm e}^{-I_0[\beta,\phi,R]}\,,
\label{eq:partaction}
\end{align}
where $I_0[\beta,\phi,R]$ is taken from 
Eq.~\eqref{eq:actionI0full}

Now, Eqs.~\eqref{eq:betaexpress}
and \eqref{eq:phiexpress} for the
stationary points can be put in
a manageable form.
For that we first define
the following quantities, 
\begin{align}
&\gamma =
\frac{16\pi^2 R^2}{(d-3)^2}
\frac{\Phi^2}{\beta^2(1-\Phi^2)^2}
\,,
\label{eq:definitions2}
\\
&\Phi = (d-3) \Omega \sqrt{\lambda} \phi\,,
\label{newPhi}
\\
&x=\frac{r_+}{R}\,,
\label{eq:definitionsx}
\\
&y=\frac{\lambda q^2}{R^{2d-6}}\,.
\label{eq:definitions}
\end{align}
So, with these definitions,
$\gamma$ substitutes the reservoir
temperature $T=\frac1\beta$,
$\Phi$ represents the electric potential $\phi$,
$x$ is the horizon radius $r_+$
in units of the reservoir radius,
and $y$ is a representation for the electric charge $q$.
Now, Eq.~\eqref{eq:phiexpress},
written in these variables,
can be inverted
to give $y = \frac{x^{2d-6}\Phi^2}{1-(1-\Phi^2)x^{d-3}}$, and 
this expression can
be used in Eq.~\eqref{eq:betaexpress},
written in the new variables,
to obtain
\begin{align}
(1-\Phi^2)x^{d-1} - x^2 + \frac{\Phi^2}{\gamma} =
0\,.
\label{eq:extremaaction}
\end{align}
Equation \eqref{eq:extremaaction}
gives the values of $x=\frac{r_+}{R}$
at the stationary points of the reduced action.
Then, by using Eq.~\eqref{eq:extremaaction} in the
expression $y = \frac{x^{2d-6}\Phi^2}{1-(1-\Phi^2)x^{d-3}}$, 
we obtain a simpler relation between $y$ and $x$, namely,
\begin{align}
&y = \gamma x^{2(d-2)}\,.
\label{eq:extrema2action}
\end{align}
Equation \eqref{eq:extrema2action} gives the values
of $y=\frac{\lambda q^2}{R^{2d-6}}$
at the stationary points of the reduced action.

\subsubsection{Analysis of the stationary points
\label{sec:Stationarypointsanalysis}}

The solutions for the Eq.~\eqref{eq:extremaaction} can be obtained
analytically for specific choices of $d$. Moreover, when $d$ is odd,
one is able to reduce by half the order of the equation. But, in general,
for 
generic dimension $d$, it is not possible to obtain an analytic
expression.

Notwithstanding, one is able to study the behavior of
Eq.~\eqref{eq:extremaaction} in terms of the parameters
$\gamma$
representing the fixed ensemble temperature, 
$\Phi$ representing the fixed ensemble
electric potential, and the dimension $d$.
One imposes that the solutions must be physical, i.e.,
the black hole must lie inside the
cavity and it must be subextremal.
The condition to lie inside the cavity is
\begin{align}
0\leq x<1\,,
\label{insidecond}
\end{align}
and the condition to be subextremal
is
$0\leq \frac{y}{x^{2(d-3)}} < 1$.
From Eq.~\eqref{eq:extrema2action}, this
latter equation can be put as 
\begin{align}
\gamma x^2 < 1\,.
\label{eq:Imposecond1}
\end{align}
Replacing Eq.~\eqref{eq:extremaaction}
in Eq.~\eqref{eq:Imposecond1},
one obtains that the condition can only be obeyed if
\begin{align}
0 \leq \Phi^2 < 1\,.
\label{Phi2cond}
\end{align}

Now,
from Eq.~\eqref{eq:extremaaction} it is useful to
define the function
\begin{align}
h(x) = (1-\Phi^2)x^{d-1} - x^2 + \frac{\Phi^2}{\gamma}
\,.
\label{hofx}
\end{align}
The values of the function at the boundary of the domain are $h(0) =
\frac{\Phi^2}{\gamma} >0$ and $h(1) = h(0)(1-\gamma)$. Note that
$\gamma$ can still be higher than 1, even though the condition 
given in
Eq.~\eqref{eq:Imposecond1} must be obeyed. Indeed, $\gamma$ is proportional 
to the temperature squared and so $\gamma$ can assume high values 
for high temperatures and fixed $\phi$ or $\Phi$.
Thus, we  separate the analysis
into
$\gamma < 1$, $\gamma = 1$, and $\gamma >1$.

\vskip 1cm
\vskip 0.6cm
\centerline{$\gamma <1$:}
\vskip 0.3cm
\noindent
For $\gamma < 1$, one has
$h(1)>0$ and so we must compute the zeros of the derivative of $h(x)$,
$h'(x)$, and the sign of the second derivative to deduce how many
zeros $h(x)$ contains. We have that
$h'(x) = x(d-1)(1-\Phi^2)\left(x^{d-3} -
\frac{2}{(d-1)(1-\Phi^2)}\right)$
and
$h''(x) = (d-1)(d-2)(1-\Phi^2)x^{d-3}-2$.
The
derivative of $h(x)$ vanishes, i.e., $h'(x_{\rm bif}) = 0$,
and the
second derivative is positive, i.e., $h''(x_{\rm bif})>0$,
at a bifurcation point
$x_{\rm bif}$
given by 
\begin{align}
x_{\rm bif} =
\left(\frac{2}{(d-1)(1-\Phi^2)}\right)^{\frac{1}{d-3}}\,.
\label{xmin}
\end{align}
The point $x_{\rm bif}$
gives the only minimum of $h(x)$, with  value
$h(x_{\rm bif})$, so that
$h(x)$ bifurcates to higher values
for $x$ lower or greater than   $x_{\rm bif}$.
In the 
case of $\gamma <1$, if the location of the minimum lies in the interval 
$x_{\rm bif}>1$, then there are no zeros of $h(x)$ since $h(x)>0$ in the
interval
$0\leq x<1$. If the minimum of $h(x)$ lies in the interval $0<x_{\rm bif}<1$, 
which implies
$
\Phi^2 <
\frac{d-3}{d-1}$,
then $h(x)$ may have zeros, but this is not a sufficient condition. 
One also must have the condition $h(x_{\rm bif})<0$, 
which implies $\gamma_{\rm bif}(\Phi,d)
\leq\gamma$,
with
\begin{align}
\gamma_{\rm bif}(\Phi,d)\equiv
\frac{(d-1)^{\frac{d-1}{d-3}}}{4^{\frac{1}{d-3}}(d-3)}\Phi^2
(1-\Phi^2)^{\frac{2}{d-3}}  \,.
\label{eq:solExcondition}
\end{align}

Therefore, in brief we have the following results. 
For
\begin{align}
\gamma <
\gamma_{\rm bif}\,,
\label{nosolutionscond}
\end{align}
there are no solutions.
For
\begin{align}
\gamma_{\rm bif}\leq\gamma <1\,,
\label{twosolutionscond}
\end{align}
there are two solutions.
We
denominate these two solutions
by $x_1$ and $x_2$, with $x_1 \leq x_2$.
Moreover,
\begin{align}
x_1 \leq
x_{\rm bif}\leq x_2\,.
\label{x1x2xmin}
\end{align}
When the equality is saturated in Eq.~\eqref{twosolutionscond},
i.e., 
$\gamma = \gamma_{\rm bif}$,
the two solutions  merge into one with $x_1=
x_{\rm bif}=x_2$. Conversely,
one can envisage
the two solutions  $x_1$ and $x_2$ as
bifurcating from $x_{\rm bif}$. 
Both solutions obey
the conditions set
in Eqs.~\eqref{insidecond}-\eqref{Phi2cond},
and also obey
\begin{align}
0\leq\Phi^2 <
\frac{d-3}{d-1}\,,
\label{stronger}
\end{align}
which is a stronger condition than
Eq.~\eqref{Phi2cond}.

\vskip 0.3cm
\centerline {   $\gamma =1$:}
\vskip 0.3cm
\noindent
For $\gamma = 1$, one of the zeros of $h(x)$ is $x=1$. We  note
however that this point
is a critical point in the sense that
it does not have the derivatives in
Eqs.~\eqref{eq:derI1}-\eqref{eq:derI2} defined. If $\Phi^2 \leq
\frac{d-3}{d-1}$, the other zero is smaller than $x=1$ or, in the case
of equality, there is no other zero. Otherwise, the other zero is
larger than $x=1$ and is nonphysical.

\vskip 0.3cm
\centerline { $\gamma >1$:}
\vskip 0.3cm
\noindent
For $\gamma >1$, the function $h(x)$ will always have one zero between
$0\leq x<1$, while the other solution will be at $x>1$, for $0<\Phi^2<1$. 
The former
zero is 
physical but we will see that it can be disregarded because of
stability, 
while the latter is unphysical since it lies outside the cavity.

\begin{figure}[b]
\centering
\includegraphics[scale=0.40]{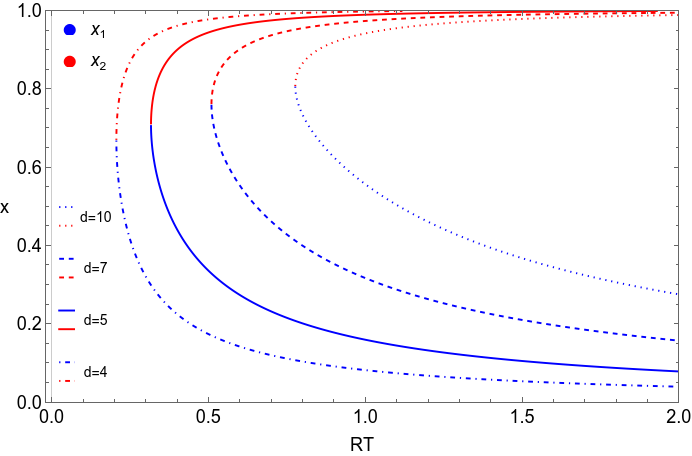}
\vskip 0.8cm
\includegraphics[scale=0.40]{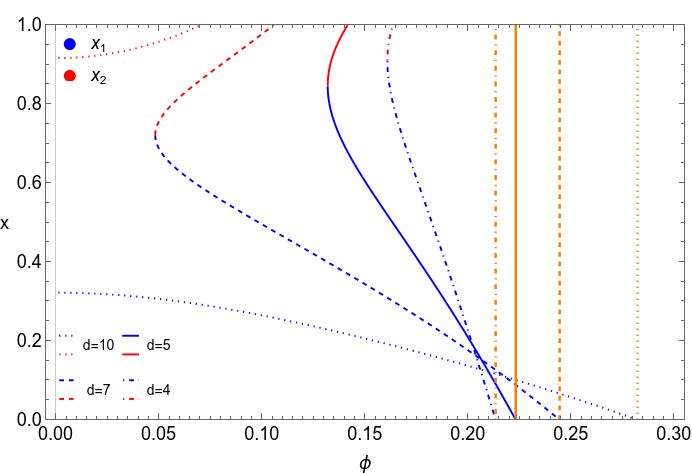}
\caption{
Top plot: Stationary points of the action, $x_1$ (in blue) and $x_2$
(in red), in function of $RT$, for $\phi = 0.02$ and for four values
of $d$: $d = 4$ in dotted lines, $d = 5$ in dashed lines, $d = 7$ in
solid lines, and $d = 10$ in dot dashed lines.
Bottom plot: Stationary points of the action, $x_1$ (in blue) and
$x_2$ (in red), in function of $\phi$, for $RT=0.3$, and the maximum
value of $\phi$ (in orange) corresponding to $\Phi=1$, for four values
of $d$: $d = 4$ in dotted lines, $d = 5$ in dashed lines, $d = 7$ in
solid lines, and $d = 10$ in dot dashed lines.
}
\label{fig:2solsphiRTconst}
\end{figure}

Note from the definition of $\gamma$ in Eq.~\eqref{eq:definitions2}
that $\gamma\propto \frac{R^2}{\beta^2}$, therefore when considering
the solutions in function of the temperature, i.e.,
in function of $\frac{R}{\beta}$ or of $RT$, 
with
$\Phi^2 < \frac{d-3}{d-1}$, the solution $x_2$ will have the range
$0\leq x_2<1$ for a finite range of $T=\frac1\beta$.
More specifically, for
$RT = \frac{d-3}{4\pi\abs{\Phi}}(1-\Phi^2)$, i.e., $\gamma=1$,
one has 
$x_2 = 1$ and so for higher values of $RT$,
one has $x_2>1$. This
behavior is not present in the
electrically uncharged case, see
\cite{Andre:2021}.
The plot of the two solutions, $x_1$ and $x_2$,
as functions of
$RT$ for constant $\phi =
0.02$ and for four values of $d$ is shown in
the top part of Fig.~\ref{fig:2solsphiRTconst}.
The plot of the two solutions, $x_1$ and $x_2$,
as functions of
$\phi$ for constant $RT =
0.3$ and four values of $d$ is shown in
the bottom part of 
Fig.~\ref{fig:2solsphiRTconst}. In these plots, the quantity 
$\phi$ was chosen instead of $\Phi$ to show the full dependence of 
the solutions in the parameter $d$, indeed $\phi$ is the quantity fixed
at the cavity while $\Phi$ is proportional to $\phi$ but the coefficient
depends on $d$.

\subsection{Perturbations around the zero loop approximation
and stability analysis}

\subsubsection{Perturbations around the zero loop approximation}

We extend the calculation of the path integral by analyzing its
expansion around the classical path.
The reduced action near the stationary points is 
\begin{align}
I_*[\beta,\phi, R; r_+,q] =
I_{0}[\beta,\phi, R]+
\sum_{ij}{I_{*0}}_{ij} \delta i \delta j
\,,
\label{eq:reducedactionperturbe}
\end{align}
where $I_{0}[\beta,\phi, R]$
is defined generically in Eq.~\eqref{eq:actionI0}
and specifically in 
Eq.~\eqref{eq:actionI0full},
and
${I_{*0}}_{ij}$ stands for 
the second derivatives of the reduced action
${I_{*}}_{ij} =
\frac{\partial^2 I_*}{\partial i \partial j}$ evaluated 
at an extremum of the action, with
$i$ and $j$ being either $r_+$ or $q$. 
Then, from Eq.~\eqref{ZasI}
we have for the path integral and the corresponding
partition function the following expression
\begin{align}
Z[\beta, \phi, R] = {\rm e}^{-I_{0}[\beta,\phi, R]} \int D[\delta
q]D[\delta r_+] {\rm e}^{-\sum_{ij}{I_{*0}}_{ij} \delta i \delta j}\,.
\end{align}
To have a proper path integral and a proper
partition function, the stationary point of
the reduced action must be a minimum.

To explicitly obtain the latter condition, we must compute the second
derivatives of the reduced action which
from Eq.~\eqref{eq:reducedaction} and
Eqs.~\eqref{eq:betaexpress}-\eqref{eq:phiexpress},
or
Eqs.~\eqref{eq:extremaaction}-\eqref{eq:extrema2action},
are
\begin{align}
   &{I_{*0}}_{r_+ r_+} = \frac{(d-2) \Omega R^{d-3} \beta}{16 \pi
   \sqrt{f} r_+^2} \mathcal{I}_{r_+r_+}\,,\label{eq:Irprp}\\
   &{I_{*0}}_{r_+ q} = \frac{(d-2) \Omega R^{d-3} \beta}{16 \pi \sqrt{f}
   r_+ q} \mathcal{I}_{r_+ q}\,,\label{eq:Irpq}\\
   &{I_{*0}}_{qq} = \frac{(d-2) \Omega R^{d-3} \beta}{16 \pi \sqrt{f}
   q^2} \mathcal{I}_{q q}\,,\label{eq:Iqq}
\end{align}
with
\begin{align}
&\mathcal{I}_{r_+ r_+} = \frac{d-3}{f
x^{2d-6}}\Big[\frac{d-3}{2}\left(x^{2d-6} - y\right)^2
\nonumber\\&-
\left(x^{2d-6}-(2d-5)y\right)
\left(1-x^{d-3}\right)\left(x^{d-3}-y\right)
\Big]\,,\label{eq:Icalpp}\\
&\mathcal{I}_{r_+ q} = -\frac{(d-3)}{x^{d-3}}\frac{\left(2 x^{d-3}
- x^{2d-6}-y\right)}{x^{d-3} - y}y\,,\label{eq:Icalpq}\\
&\mathcal{I}_{qq} = 2\frac{1-x^{d-3}}{x^{d-3} -
y}y\,.\label{eq:Icalqq}
\end{align}
The matrix ${I_{*0}}_{ij}$ is positive definite if the pivots of the
matrix after Gauss elimination are positive or, in this case, since
the matrix has rank 2, the first element on the diagonal and the
determinant need to be positive, i.e.,
\begin{align}
&\mathcal{I}_{r_+ r_+} > 0\,,\label{eq:condstab1}\\
&\mathcal{I}_{r_+ r_+}\mathcal{I}_{q q}-\mathcal{I}_{r_+
q}^2>0\,.
\label{eq:condstab2}
\end{align}
Since from Eq.~\eqref{eq:Icalqq}
$\mathcal{I}_{q q}$ is always positive, we have that
the
last condition, Eq.~\eqref{eq:condstab2}, is sufficient. Using
Eqs.~\eqref{eq:Icalpp}-\eqref{eq:Icalqq}
and~\eqref{eq:extrema2action}, we have that the condition in
Eq.~\eqref{eq:condstab2} reduces to
\begin{align}
  - (d-3)\gamma x^{d-1}+  (d-1)x^{d-3}  - 2 > 0\,.
\label{eq:stabilityx}
\end{align}
This is the condition for a stationary point, given by
Eqs.~\eqref{eq:extremaaction} and \eqref{eq:extrema2action}, to be a
minimum of the action.

\subsubsection{Stability analysis}

Since the stability condition 
is applied to the stationary points of the reduced action, 
one can use Eq.~\eqref{eq:extremaaction} to simplify the condition
given in 
Eq.~\eqref{eq:stabilityx}
in the following way. One can 
rewrite Eq.~\eqref{eq:extremaaction} to get 
$\gamma = \frac{\Phi^2}{x^2 - (1-\Phi^2)x^{d-1}}$ and substitute 
$\gamma$ into Eq.~\eqref{eq:stabilityx}
obtaining a condition depending 
solely on $r_+$ and $\Phi$. So, on using 
Eq.~\eqref{eq:extremaaction}, one gets that
Eq.~\eqref{eq:stabilityx} indeed simplifies to
\begin{align}
 \frac{((d-1)(1- \Phi^2)x^{d-3} - 2) (1 - x^{d-3})}
 {1 - (1-\Phi^2)x^{d-3}}
 > 0\,\,\,.\label{eq:stabilitysimple}
\end{align}
The physical range of solutions is 
$0\leq x^{d-3}<1$, and so the denominator is always greater
than zero.
Therefore, there is thermodynamic stability if
the solution satisfies 
\begin{align}
x >  x_{\rm bif} \,,\label{eq:criterionstability}
\end{align}
where
$x_{\rm bif} = \left(\frac{2}{(d-1)(1-\Phi^2)}
\right)^{\frac{1}{d-3}}$, see Eq.~\eqref{xmin}.
Note that $x_{\rm bif}$ is the value of $x=\frac{r_+}{R}$ from which
the two solutions $x_1$ and $x_2$ bifurcate from, and so one always has
that $x_1 \leq x_{\rm bif}\leq x_2$.
Thus, the bifurcation radius is equal to the marginal
thermodynamic stability
radius.
In the electrically uncharged case
the bifurcation and marginal thermodynamic stability
radius and the photon sphere radius coincide,
so it is worth to see if this stands in the
electrically charged case, see
Appendix~\ref{photonandbhcdahlandreasson}.
We now analyze the stability for each case of $\gamma$.

\vskip 0.3cm
\centerline {$\gamma <1$:}
\vskip 0.3cm
\noindent
For the case of $\gamma_{\rm bif}<\gamma <1$ and $\Phi^2 <
\frac{d-3}{d-1}$, one has $x_1 < x_{\rm bif}< x_2 < 1$, so $x_1$ is
unstable and it corresponds to a saddle point of the action, while
$x_2$ is stable and corresponds to a minimum of the action. For the
case of $\frac{d-3}{d-1}\leq \Phi^2<1$, both solutions $x_1$ and $x_2$ are
not physical as they lie outside the cavity.  We now analyze the case
$x_1=x_{\rm bif}=x_2$.  One must note that $x_{\rm bif}$ is the only
solution of Eq.~\eqref{eq:extremaaction} for
$\gamma = \gamma_{\rm bif}$ and
for which Eq.~\eqref{eq:stabilitysimple} is an equality
rather than an inequality. Thus, one cannot specify the critical
point with only second derivatives of the action, perhaps third
derivatives will do.  However, by inspection of the
action $I(x,y)$ at this point one finds that it is a saddle point.

\vskip 1cm
\vskip 0.3cm
\centerline {$\gamma = 1$:}
\vskip 0.3cm
\noindent
For the case of $\gamma =1$ and $\Phi^2 < \frac{d-3}{d-1}$, the
solution $x_1$ is unstable, while the solution $x_2$ reaches the
boundary of the cavity, $x_2 = 1$, and at this point the derivatives
of the action are not well defined, so one cannot specify the
stability. For the case of $\Phi^2=\frac{d-3}{d-1}$
the two solutions $x_1$ and $x_2$ reach the boundary and
 the derivatives
of the action are not well defined as well.
For the case of $\frac{d-3}{d-1}<\Phi^2 <1$, the solution
$x_2$ lies outside the cavity, thus not physical, while $x_1$ is at
the boundary of the cavity, $x_1 = 1$, and one cannot specify its
stability.

\vskip 0.5cm
\centerline {$\gamma >1$:}
\vskip 0.3cm
\noindent
For the case of $\gamma > 1$ and for $0< \Phi^2 <1$, the solution
$x_1$ resides inside the cavity and it is still unstable, while the
solution $x_2$ is not physical as it is outside the cavity, i.e., $x_2
> 1$.

\subsection{Most probable configurations}

From Eq.~\eqref{ZasI}, we see that the paths with less $I_*$, or from
Eq.~\eqref{eq:partaction} the paths with less $I_0$, are the ones that
contribute more to the partition function, and so yield the most
probable states.  Here we deal with stable solutions and among those
we want to find the one that gives the most probable state.

In the electrically uncharged case done in
\cite{York:1986} for $d=4$ and
in \cite{Andre:2020,Andre:2021}
for generic $d$, the comparison between the
stable black hole solution and hot flat space was made.  The stable
black hole is a stationary point of the reduced action, and the hot flat
space solution is an extra stationary point.  The most probable state is
the one with the lowest value of the action.  In the case of the
uncharged black
holes, the value of the action $I_0$ depends on $\beta$, while in the
case of hot flat space one has $I_{\rm hot\, flat\, space} = 0$.  In
\cite{Andre:2020,Andre:2021}, it was shown
for any dimension $d\geq4$ that $I_0 < I_{\rm hot\, flat\, space}$
if $\beta$ is such that $\frac{r_+}{R} > \frac{r_{\rm Buch}}{R}$,
where $r_{\rm Buch}$ is the Buchdahl radius. In
addition, in
\cite{Andre:2020,Andre:2021}, a comparison between the stable black
hole solution and quantum hot flat space was also done.

In the electrically charged case, one can also make a comparison of
the stable black hole with the charged equivalent of the uncharged hot
flat space, as we will see.  The electrically charged case is more
rich than the uncharged one.  In the charged case, besides the
stationary point related to the stable black hole, there are two
critical points that are possible stable solutions of the ensemble.
One critical point,
which is a stationary point indeed,
is $r_+ = 0$ and $q = 0$, corresponding
to a cavity without a black hole and without charge.  This seems
unphysical for a fixed nonzero value of $\phi$, since it means that
there is a difference of electric potential, which in turn implies the
existence of an electric field and thus of an electric charge.  For
this reason, $q=0$ seems unphysical.  However, we have to recall that
the path integral approach in the semiclassical approximation deals
intrinsically with quantum systems, and when one writes $q=0$, one
should mean $q$ of the order of the Planck charge, and a particle,
say, carrying such a charge should be envisaged as having the
dimensions of the order of a Planck length or a bit higher.  Thus, we
have to seek a corresponding action for such a particle in a reservoir
of fixed $R$ and $\beta$.  The other  critical point is $r_+=R$
and $\sqrt{\lambda} q=R^{d-3}$, so that $r_+= (\sqrt{\lambda}
q)^\frac1{d-3}=R$.  This critical point corresponds to an extremal
black hole with the horizon localized at the radius of the cavity,
meaning that the volume of the Euclidean space is zero, which
can require a different procedure.
However, again, this is a quantum system treated
semiclassically, and so one should think of a black hole almost at its
extremal state, failing to be extremal by a Planck charge and not
touching the reservoir at $R$ by a Planck length. Thus, in this state
the approach is still valid under our treatment, and we have to find
the value of the action for a large extreme black hole in a reservoir
with fixed $R$ and $\beta$.  So, the question of whether the stable
black hole is the ground state or there is another ground state to
which the black hole can make a transition is a pertinent one.

Let us now deal with the first  critical point $r_+ = 0$ and $q =
0$, at which the derivative of the action in order to $q$ is not
well defined. Nonetheless,
one can argue that this critical point can be considered 
as a local minimum of the action in the physical domain, see
Appendix~\ref{app:gradactioncavity} for the calculation. In an attempt
to describe an equivalent of hot flat space, we consider a hot sphere,
made of a perfect conductor material, with a certain radius $r_{\rm hs}$,
inside the reservoir at constant $\beta$ and $\phi$, and with its
center situated at the center of the reservoir. There is no
gravitational interaction, i.e., the constant of gravitation is put to
zero. This is equivalent to consider only the Maxwell term in
Eq.~\eqref{eq:EucAction1}. One must then consider a fixed
radius $r_{\rm hs}$ for
the hot sphere
conductor, in the boundary conditions. From the Gauss constraint,
the charge of the conducting sphere can be related directly to the
value of $\phi$.  Indeed, one has $\phi =
\frac{q}{(d-3)\Omega}\left(\frac{1}{r_{\rm hs}^{d-3}}-
\frac{1}{R^{d-3}}\right)$, see also Eq.~\eqref{eq:phiexpress}.
Therefore, the action for this cavity being $I=-\frac12 q\beta\phi$,
turns, for the perfect conducting hot sphere in flat space, into the
expression
\begin{align}
I_{\rm hot \,sphere} = - \frac12\, \frac{(d-3)\Omega}
{\frac{1}{r_{\rm hs}^{d-3}}-
\frac{1}{R^{d-3}}}\beta\phi^2
\,.
\label{Ihotflatsphere}
\end{align}
One can then compare the action of the conducting hot sphere
with radius $r_{\rm hs}$ given in
Eq.~\eqref{Ihotflatsphere} with the action of the stable configuration
of the charged black hole, which is Eq.~\eqref{eq:reducedaction} with
the largest positive solution of Eq.~\eqref{eq:extremaaction},
i.e., the $r_{+2}$ solution.  From
Eq.~\eqref{Ihotflatsphere}, it is clear that if $r_{\rm hs}$ is high, of the
order of $R$, say, then $I_{\rm hot \,sphere}$ is large and
negative and so the hot flat sphere is
the most probable solution when
compared to the stable black hole $r_{+2}$. On the other hand, if
$r_{\rm hs}$ is tiny, as we expect to be when dealing with a case analogous
to hot flat space, then $I_{\rm hot \,sphere} =0$, or
approximately zero.  In this situation one can say that $I_{\rm hot
\,sphere}$ is indeed $I_{\rm hot \,flat\,space}$ which is a
configuration with zero action. The stable black hole does have
positive action
for low temperatures $T$, specifically, 
near the minimum temperature where the stable black
hole exists.  Therefore, one finds that the tiny charged sphere that
emulates hot flat space is more probable for a small interval of
low temperatures when compared with the stable black hole.  Conversely,
the black hole is more probable for a large interval of temperatures,
in fact, when the solution of the stable black hole has
\begin{align}
\frac{r_{+2}^{d-3}}{R^{d-3}} \geq \mu m + \sqrt{\mu^2
m^2 - \lambda q^2}
\,.
\label{conditionbhmoreprobable}
\end{align}
where $\mu = \frac{8\pi}{(d-2)\Omega}$, see Eq.~\eqref{eq:mu}, and
also $\mu m R^{3-d}= -\frac{4 (d-2)^2}{(d-1)^2(d-3)^2} + \frac{2(d-2)((d-2)^2
+ 1)}{(d-1)^2(d-3)^2}\sqrt{1 + \frac{(d-1)^2 (d-3)^2}{4(d-2)^2}
\frac{\lambda q^2}{R^{2d-6}}}$,
see Appendix \ref{photonandbhcdahlandreasson} for this equality.
When Eq.~\eqref{conditionbhmoreprobable}
is obeyed then the action for the black hole $r_{+2}$
is negative
and the black hole is more
probable, when Eq.~\eqref{conditionbhmoreprobable}
is not obeyed the tiny charged sphere
is more probable. Notice that this radius $R$ does not have a
connection to the Buchdahl-Andre\'asson-Wright radius~\cite{Wright:2015},
a radius that generalizes the Buchdahl bound
for $d$-dimensional
self-gravitating  electric charged spheres. 
The horizon radius with
zero action is equal or lower than the Buchdahl-Andre\'asson-Wright radius
in the case of $d=4$,
with a difference up to $0.004$ in $\frac{\mu m}{R}$, and being equal
in the uncharged case and the extreme case $\sqrt{\lambda}q = R$.
Thus, the equality in the uncharged situation
of the minimum most probable
radius of a black hole in the canonical ensemble and
the Buchdahl radius does not hold when other fields
are added. So, it is a very restricted equality
holding only in the pure gravitational situation.

Let us now deal with the second critical point $r_+=R$ and
$\sqrt{\lambda}q=R^{d-3}$, i.e., an extremal black hole with the
horizon localized at the radius of the cavity, bearing in mind that
the precise extremality and the precise location can fluctuate by
Planck order quantities. We note that this is a critical point in the
sense that the gradient of the action is not defined. Indeed, one can
calculate the gradient of the reduced action in
Eq.~\eqref{eq:reducedaction} and make the limit to $r_+=
(\sqrt{\lambda}q)^{\frac1{d-3}}=R$ along the curve $\frac{r_+}{R} = (1
- \epsilon)^{\frac1{d-3}}$ and $\frac{\sqrt{\lambda} q}{R^{d-3}} =
\sqrt{(1 - \eta \epsilon)}$, where $\eta$ is a positive constant and
$\epsilon$ parameterizes the curve.  The constant $\eta$ is restricted
to the physical domain of the action, with the condition $\eta >
2$. After substituting the variables by the parameterization of the
curve in the expression of the gradient and performing the limit
$\epsilon \rightarrow 0^+$, the gradient assumes
an expression 
that depends on the constant $\eta$.
Since the limit is different for different values of $\eta$, then the
gradient cannot be defined in that point,
but one can still analyze
the directional derivatives along the considered paths. The
directional derivatives along decreasing $\epsilon$, i.e., from lower
$r_+$ and $q$ toward  $r_+=
(\sqrt{\lambda}q)^{\frac1{d-3}}=R$, may be either positive, zero, or
negative,
and so the critical point does not resemble a local minimum.
In particular, there is a set of temperatures and electric potential
given by the condition $\gamma = 1$, where the stable black hole
solution tends to this extremal black hole. Indeed, it can be seen
that for such values of temperature and electric potential, there is a
value of $\eta$ in which the limit of the gradient vanishes, but the
fact still remains that the gradient is undefined here, see
Appendix~\ref{app:gradactioncavity} for a detailed
analysis
of the gradient at this critical point. 
Nevertheless, this critical point may be
smoothed up by taking in consideration higher loops in the path
integral or a different theory of gravity.  The action for this
critical point can be taken from Eq.~\eqref{eq:reducedaction}, i.e.,
$I_{\rm extreme\,black\, hole} = \frac{(d-2)\Omega
R^{d-3}\beta}{8\pi}\Bigl(1- \Bigr.  \left.\sqrt{f(R,r_+,q)}\right) -q
\beta \phi - \frac{\Omega r_+^{d-2}}{4}$, where $f(R,r_+,q)$ is taken
from Eq.~\eqref{fd} with $r_+$ and $q$ having extremal values, so that
$R=r_+$ and $f(R,r_+,q)=0$.  Then,
\begin{align}
I_{\rm extreme\,black\, hole} =& \frac{(d-2)\Omega
R^{d-3}\beta}{8\pi}
\nonumber
\\
&
-\frac{R^{d-3}}{\sqrt\lambda} \beta \phi -
\frac{\Omega R^{d-2}}{4}\,.
\label{Icoldcurvedspace}
\end{align}
So $I_{\rm extreme\,black\, hole}$ has to be analyzed for each $R$,
$\beta$ and $\phi$, and compared with the action for the stable black hole
$r_{+2}$.  It seems that the stable black hole is
always a more probable
configuration than the extreme black hole with horizon at the cavity.

\section{Thermodynamics of
the $d$-dimensional Reissner-Nordstr\"om
black hole space in a cavity}
\label{sec:GrandCanonical}

\subsection{Connection between the grand canonical potential $W$ of
thermodynamics and the statistical mechanics grand canonical partition
function $Z$ found from the path integral action $I$}

We can relate the partition function $Z$ of the grand canonical ensemble
directly to the thermodynamic grand potential $W$
and therefore obtain the
thermodynamics of the system.  The relation is
\begin{align}
Z(\beta,\phi,\bm{\gamma}) = {\rm e}^{-\beta W[\beta, \phi,
\bm{\gamma}]}\,, 
\label{eq:PartGrandCanonical} 
\end{align}
or $\beta W=-\ln Z$. 

In the semiclassical approximation, one has 
$Z={\rm e}^{-I_0}$
and so 
one has
$\beta W[\beta,\phi,R] = I_0[\beta,\phi,R]$.
By taking in consideration that $\beta=\frac1T$, one
can write the relation between the
thermodynamic grand potential $W$
and the action $I_0$ as
\begin{align}
W[T,\phi,A(R)] = T\,I_0[T,\phi,R]\,.
\label{grandpot}
\end{align}
Therefore, from Eqs.~\eqref{eq:actionI0full}
and \eqref{grandpot}, the grand potential
$W$ can be written as
\begin{align}
W=& \frac{(d-2)\Omega
R^{d-3}}{8\pi}
\Bigl(1-\sqrt{f\left( R,T,\phi\right)}\Bigr)\nonumber \\
&-
T\frac{\Omega r_+^{d-2}(R,T,\phi)}{4}-
q(R,T,\phi) \phi \,,
\label{eq:grandpotentialactionI0full}
\end{align}
where $f(R,T,\phi)$
can be taken from Eq.~\eqref{fd} and the
solutions of Eqs.~\eqref{eq:betaexpress}
and \eqref{eq:phiexpress}.

The grand canonical potential
$W$ can be written by definition 
as a Legendre transformation of the
mean energy $E$.
Written in terms of $E$,  the
mean electrical charge $Q$,
the electric potential $\phi$, the entropy $S$, and
the temperature $T$, one has
\begin{align}
W = E - TS-Q \phi\,,
\label{grandpotthermodef}
\end{align}
where $E=E(S,Q,A)$. Moreover, the entropy,
the mean charge, and the
thermodynamic pressure are obtained 
from the derivatives of the grand potential, which
then allows one to find the mean energy by 
Eq.~\eqref{grandpotthermodef}.

The connection between the statistical path integral and
thermodynamics can be summarized by Eq.~\eqref{grandpot}, in the zero
loop approximation. However, it is interesting to see the relationship
between the choice of the path with minimum action, in the path
integral, and the thermodynamics of the system, in the zero loop
approximation. Indeed, it seems that if one identifies the expression
of the quantities $E$, $S$ and $Q$ in the reduced action, the choice
of the path with minimum action imposes that the temperature and the
electric potential are partial derivatives of the energy, as one would
expect from the first law of thermodynamics, see
Appendix~\ref{canthermo}.

\subsection{Thermodynamic quantities and relations}

\subsubsection{Mean energy, entropy, mean charge,  and
thermodynamic pressure}

The grand potential $W=W[T,\phi,R]$
of the cavity is given by 
Eq.~\eqref{eq:grandpotentialactionI0full}.
Since $A=4\pi R^2$, see Eq.~\eqref{areaA},
one can trade $R$ for $A$ in $W$ to have
the dependence $W=W[T,\phi,A]$.
One can then write the differential 
of this form of the grand potential $W$ as
\begin{align}
dW = -SdT
-
pdA
-
Qd\phi\,,
\label{grandpotdiffform}
\end{align}
where from Eqs.~\eqref{grandpotthermodef}
and \eqref{grandpotdiffform} one has
$S=-\left(\frac{\partial W}{\partial T}\right)_{A,\phi}$,
$p\equiv-\left(\frac{\partial E}{\partial A}\right)_{S,Q}=
-\left(\frac{\partial W}{\partial A}\right)_{T,\phi}$,
and $Q=-\left(\frac{\partial W}{\partial \phi}\right)_{A,T}$.
Here,
a quantity in a subscript means that the partial derivative
is performed with the corresponding quantity being kept
constant, e.g., $\left(\frac{\partial W}{\partial T}\right)_{A,\phi}$
means partial derivative of $W$ in relation to $T$ with
$A$ and $\phi$ kept constant.
In order to obtain the thermodynamic quantities, we must evaluate 
the derivatives just given of the grand potential $W$.

We  start by calculating
the entropy $S=- \left(\frac{\partial W}{\partial T}\right)_{A,\phi}$.
From Eq.~\eqref{eq:grandpotentialactionI0full} we see that
$W=W(T,\phi,A,r_+(T,\phi,R), q(T,\phi,R))$. So, using the chain rule
one has
$S=-
\left(\frac{\partial W}{\partial T}\right)_{A,\phi}=
-
\left(\frac{\partial W}{\partial T}
\right)_{r_+,q,A,\phi}
-\left(\frac{\partial W}{\partial r_+}
\right)_{q,T,A,\phi}
\left(\frac{\partial r_+}{\partial T}
\right)
-
\left(\frac{\partial W}{\partial q}
\right)_{r_+,T,A,\phi}
\left(\frac{\partial q}{\partial T}
\right)$.
Now, the black hole solutions given by $r_+(T,\phi,R)$ and
$q(T,\phi,R)$ obey by definition the conditions $\left(\frac{\partial
W}{\partial r_+}\right)_{q,T,A,\phi} \equiv 0$ and $\left(\frac{\partial
W}{\partial q}\right)_{r_+,T,A,\phi} \equiv 0$, yielding thus simply $S=
-
\left(\frac{\partial W}{\partial T}\right)_{A,\phi}=
-
\left(\frac{\partial W}{\partial T}\right)_{r_+,q,A,\phi}$.  So,
Eq.~\eqref{eq:grandpotentialactionI0full} yields directly
\begin{align}
S = \frac{A_+}{4}\,,
\label{eq:entropyrp}
\end{align}
with $A_+$ is the area of the horizon given by
$A_+=\Omega r_+^{d-2}$. Thus, Eq.~\eqref{eq:entropyrp}
yields that the black hole entropy is indeed
given by
the Bekenstein-Hawking entropy formula.

In the
same manner, one can calculate the electric charge
$Q$ to give the expression
$Q = - \left(\frac{\partial W}{\partial \phi}\right)_{T,A}
= - \left(\frac{\partial W}{\partial \phi}\right)_{r_+,q,T,A} 
$, yielding
\begin{align}
Q =  q\,,
\label{eq:meanQ}
\end{align}
so the thermodynamic value of the electric charge $Q$
is equal to the typical electric
charge $q$ of a Reissner-Nordstr\"om black hole.

The pressure
is given by
$p=
- 
\left(\frac{\partial W}{\partial A}\right)_{T,\phi}
=
- \frac{1}{(d-2)\Omega R^{d-3}}
\left(\frac{\partial W}{\partial R}\right)_{r_+,q,T,\phi}$
and so it has the form
\begin{align}
p= \frac{d-3}{16\pi R \sqrt{f}}\left(\left(1 -
\sqrt{f}\right)^2 - \frac{\lambda q^2}{R^{2d-6}}\right)\,,
\label{eq:meanpressure}
\end{align}
which is the gravitational tangential pressure at
the reservoir at radius $R$.

Finally, one can calculate the energy by
putting Eqs.~\eqref{eq:entropyrp}-\eqref{eq:meanpressure} into
Eq.~\eqref{grandpotthermodef} and finding that
\begin{align}
E = \frac{(d-2)\Omega R^{d-3}}{8\pi} \left(1 -
\sqrt{f} \right)\,,
\label{eq:Energy}
\end{align}
which is the thermodynamic energy,
a quasilocal energy, evaluated at 
radius $R$.

One can  verify from the previous equations
that 
\begin{align}
TdS=   dE  +pdA - \phi dQ\,,
\label{1stlawusual}
\end{align}
i.e., the first law of thermodynamics
for the system holds. In Eq.~\eqref{1stlawusual},
$T=\frac1\beta$ can be taken from 
Eq.~\eqref{eq:betaexpress},
$S$ is given by Eq.~\eqref{eq:entropyrp},
$E$ is given by
Eq.~\eqref{eq:Energy}, $p$, is given by
Eq.~\eqref{eq:meanpressure},
 $\phi$
is given by Eq.~\eqref{eq:phiexpress}, and 
$Q$ is given by
Eq.~\eqref{eq:meanQ}.

It
is worth pointing out that the expressions of the entropy, the mean
charge, and the energy can be taken naively from the comparison of
Eq.~\eqref{eq:grandpotentialactionI0full} with
Eq.~\eqref{grandpotthermodef}, but this is in fact true
only because by definition of $I_0$, and thus of $W=TI_0$,
the conditions $\left(\frac{\partial
W}{\partial r_+}\right)_{q,T,A,\phi} \equiv 0$
and $\left(\frac{\partial
W}{\partial q}\right)_{r_+,T,A,\phi} \equiv 0$,
are identically satisfied. Thus, the extrema of the
reduced action $I_*$
can be interpreted as
points that set the first law of
thermodynamics in Eq.~\eqref{1stlawusual}, see
also Appendix~\ref{canthermo}.

\subsubsection{Euler equation, Gibbs\hskip0.04cm-Duhem relation,
and Smarr formula}

With the thermodynamic variables
$E$, $A$, and $Q$
determined, one is able to obtain the Euler equation and the 
Gibbs-Duhem relation, i.e., the energy in terms of the remaining
thermodynamic variables and differential relation between 
the thermodynamic variables, respectively. 

The energy in Eq.~\eqref{eq:Energy} can be
rewritten in terms of the entropy $S$, surface area of the cavity $A$,
and charge $Q$ as
\begin{align}
E =&
\frac{(d-2)A^{\frac{d-3}{d-2}}
\Omega^{\frac{1}{d-2}}}{8\pi}\nonumber\\&
\left(1-\sqrt{\left(1-\left(\frac{4
S}{A}\right)^{\frac{d-3}{d-2}}\right)\left(1-\frac{\lambda Q^2
\Omega^{2\frac{d-3}{d-2}}}{(4SA)^{\frac{d-3}{d-2}}}\right)}
\right)\,.\label{eq:energyintermsof}
\end{align}
One can then use the Euler's homogeneous function theorem considering
that under a rescaling $\nu$  of its arguments, the energy as a function has
the property that $E\left(\nu S,\nu A,\nu Q^{\frac{d-2}{d-3}}\right) =
\nu^{\frac{d-3}{d-2}} E\left(S,A, Q^{\frac{d-2}{d-3}}\right)$. We thus
have an integrated version of the first law of thermodynamics given by
\begin{align}
E = \frac{d-2}{d-3}(TS - pA) + \phi
Q\,,\label{eq:1stlawintegrated}
\end{align}
which is the Euler equation for
the system of a
$d$-dimensional electrically charged black holes in a
heat reservoir.
By differentiating Eq.~\eqref{eq:1stlawintegrated} and considering
that $dE = TdS - pdA + \phi dQ$, one obtains
\begin{align}
  TdS - pdA + (d-2)(SdT-Adp) + (d-3) Qd\phi = 0\,.
\label{eq:gibbs}
\end{align}
which is the Gibbs\hskip0.04cm-Duhem relation for
the system of a
$d$-dimensional electrically charged black hole in a
heat reservoir.

In the limit of infinite radius of the cavity, one 
can obtain the 
Smarr formula from Eq.~\eqref{eq:1stlawintegrated}. 
Indeed, in the limit of infinite radius $R$,
the temperature in Eq.~\eqref{eq:betaexpress} 
reduces to the Hawking temperature, i.e., $T=T_{\rm H}=
\frac{d-3}{4\pi}\left(\frac{1}{r_+} 
- \frac{\lambda q^2}{r_+^{2d-5}} \right)$,
the electric
potential in Eq.~\eqref{eq:phiexpress} reduces to the
electric potential of the Reissner-Nordstr\"om black hole, 
i.e., $\phi = \phi_{\rm H} = \frac{q}{(d-3)\Omega r_+^{d-3}}$,
 the quantity $p A$ with $p$ in
Eq.~\eqref{eq:meanpressure} being proportional to $\frac{1}{R^{d-3}}$
vanishes, and
the 
energy of the system in Eq.~\eqref{eq:Energy} reduces 
to the ADM mass, i.e.,
$E = m$,
Therefore, from 
Eq.~\eqref{eq:1stlawintegrated} and from the considerations
above, the Smarr formula is given by 
\begin{align}
m = \frac{d-2}{d-3} T_{\rm H}S + \phi_{\rm H}
Q \,.\label{eq:smarr}
\end{align}
The Smarr formula of Eq.~\eqref{eq:smarr} can only be valid
for the small black hole solution of the grand
canonical ensemble, 
since it is the only solution that exists in this limit.

\subsection{Stability in terms of thermodynamic variables}
\label{sec:Equilibria}

In a thermodynamic system with fixed size, fixed temperature, and fixed
electric potential, attached to a heat reservoir,
energy, entropy, and electric
charge can flow from the system to the reservoir and back.
In any thermodynamic process in such a system,
the grand canonical potential $W$ tends to
decrease down to its minimum or stay at its minimum.  In particular,
a spontaneous process in the
grand canonical ensemble can never increase the grand canonical
potential $W$.

To see this one must resort to the second law
of thermodynamics applied to the total structure.
Indeed, a variation $dS$ in entropy
in the system, plus a variation $dS_{\rm reservoir}$
in entropy of the reservoir, add to
a variation $dS_{\rm total}$ of the 
total entropy of the system plus reservoir, 
$dS_{\rm total}=dS+dS_{\rm reservoir}$.
Suppose the thermodynamic
system  absorbs energy $dE$
and charge $dQ$ from the reservoir.
Then the reservoir
absorbs  energy $-dE$ and charge $-dQ$.
The first law
of thermodynamics then states
that the change in entropy in the reservoir is
$TdS_{\rm reservoir}=
-dE+\phi dQ$, where 
since
the reservoir remains in internal equilibrium
by definition,
its temperature and its electric
potential are kept constant,
i.e., have the original values $T$ and $\phi$.
So, we can write for the total change in entropy, 
$TdS_{\rm total}=TdS-dE+\phi dQ=-d(E-TS-\phi Q)=
-d{\bar W}$,
where we used that $T$ and $\phi$ are
constant since they are reservoir values.
We have also defined ${\bar W}$ as
\begin{align}
{\bar W}[{\bar T},A,{\bar \phi}]\equiv E({\bar T},A,{\bar \phi})
- T S({\bar T},A,{\bar \phi}) - \phi Q({\bar T},A,{\bar \phi})\,,
\label{eq:Wperturb}
\end{align}
as the grand canonical potential related
to the nonequilibrium situation.
Note that due to the variation to a 
 nonequilibrium situation the thermodynamic
 system has in general 
a new temperature
$\bar T$ and a new potential $\bar \phi$
different from $T$ and $\phi$ of the reservoir.
The new quantities that arise in the variation
of the nonequilibrium situation,
$E$, $S$, and $Q$, have
the same functional form of ${\bar T}$, $A$, and 
${\bar \phi}$, as they had of $T$, $A$, and $\phi$
before the nonequilibrium process set in,
but ${\bar W}[{\bar T},A,{\bar \phi}]$
has a different functional form,
since $T$ and $\phi$
that appear in Eq.~\eqref{eq:Wperturb}
are quantities of the heat reservoir
fixed 
by assumption at the origin.
The area $A$ has been kept fixed in the process. 
Thus, returning to the total entropy,
one has in brief $TdS_{\rm total}=
-d{\bar W}$. Since mandatorily $dS_{\rm total}\geq0$
by the second law, one deduces
$d{\bar W}\leq0$. Any spontaneous process decreases
the grand canonical potential. For
further discussion on these issues see
Sec.~8.2 and the following sections of \cite{reif}.

The equilibrium is assured if
\begin{align}
{\bar T}=T\,,\quad\quad {\bar \phi}=\phi\,,
\label{eq:WS0equilibrium}
\end{align}
in which case ${\bar W}$ is at its minimum. 
To be stable, 
fluctuations 
in the temperature $T$ of the thermodynamic system
should tend to increase $W$,
fluctuations in the electric potential $\phi$
should tend to increase $W$, and a mixture of the
two fluctuations
should tend to increase $W$.
So, stability is assured if
\begin{align}
\left(\frac{\partial^2 {\bar W}}{\partial {\bar T}^2}
\right)_{{\bar \phi},A} > 0\,,
\label{eq:WS2}
\end{align}
\begin{align}
\left(\frac{\partial^2 {\bar W}}{\partial {\bar T}^2}\right)_{{\bar \phi},A} 
\left(\frac{\partial^2 {\bar W}}{\partial {\bar \phi}^2}\right)_{{\bar T},A} 
-
\left(\frac{\partial^2 {\bar W}}{\partial {\bar T} \partial {\bar \phi}}
\right)_{{\bar T};{\bar \phi},A} ^2
> 0\,,\label{eq:WSQ}
\end{align}
\begin{align}
\left(\frac{\partial^2 {\bar W}}{\partial {\bar \phi}^2}
\right)_{{\bar T},A} > 0
\,,
\label{eq:WQ2}
\end{align}
where in the second term of
Eq.~\eqref{eq:WSQ} the cross derivatives
are performed maintaining constant the quantity
that is not being differentiated,
and  all 
the derivatives are to be
calculated at the solutions of the ensemble.
Additionally, only two conditions from
Eqs.~\eqref{eq:WS2}-\eqref{eq:WQ2} are sufficient, we choose
Eqs.~\eqref{eq:WS2} and~\eqref{eq:WSQ}.
From the expression of ${\bar W}$,
one can show that
$\left(\frac{\partial^2 {\bar W}}{\partial {\bar T^2}}\right)_{A,
{\bar \phi}} =
\left(\frac{\partial  S}{\partial  T}\right)_{A,\phi}$,
where the bars have been dropped
on the right-hand side of the equality because $S$
 has
the same functional form of ${\bar T}$, $A$, and 
${\bar \phi}$, as it has of $T$, $A$, and $\phi$,
and at equilibrium ${\bar T}=T$ and the area
was never perturbed.
In the same manner, 
$\left(\frac{\partial^2 {\bar W}}{\partial {\bar \phi}^2}
\right)_{{\bar T},A} =\left(\frac{\partial Q}{\partial
\phi}\right)_{T,A}$,
and
$\left(\frac{\partial^2 {\bar W}}{\partial {\bar T}
\partial {\bar \phi}}\right)_{{\bar T};{\bar \phi},A}
=\left(\frac{\partial Q}{\partial T}\right)_{A,\phi}=
\left(\frac{\partial S}{\partial \phi}\right)_{T,A}$.
The  two
sufficient conditions,
Eqs.~\eqref{eq:WS2} and~\eqref{eq:WSQ},
can then be written as
\begin{align}
& \left(\frac{\partial S}{\partial T}\right)_{A,\phi} > 0\,,
\label{eq:WS2better}\\
& \left(\frac{\partial Q}{\partial\phi} \right)_{T,A}
\left(\frac{\partial S}{\partial T} \right)_{A,\phi} 
- \left(\frac{\partial S}{\partial \phi}\right)^2_{T,A} > 0\,
\label{eq:WSQbetter}
\end{align}
respectively.

We now define two quantities. First,
we define the isochoric heat capacity 
at constant electric potential as
\begin{align}
C_{A,\phi} = 
T\left(\frac{\partial S}{\partial T}\right)_{A,\phi}\,.
\label{CAphi}
\end{align}
Second, we define
the adiabatic electric susceptibility as 
$\chi_{S,A}
= \left(\frac{\partial
Q}{\partial \phi}\right)_{S,A}$. It can
also be written in terms 
of the derivatives present in Eq.~\eqref{eq:WSQbetter}
from a change of variables 
$Q(T,A,\phi)$ to $Q(T(S,A,\phi), A,\phi)$, where $T(S,A,\phi)$ is 
the inverse function of $S(T,A,\phi)$. Indeed, one gets
\begin{align}
\chi_{S,A}=
\frac{\left(\frac{\partial Q}{\partial\phi} \right)_{T,A} 
\left(\frac{\partial S}{\partial T}\right)_{A,\phi}
- \left(\frac{\partial S}{\partial \phi}\right)^2_{T,A} }
{\left(\frac{\partial S}{\partial T}\right)_{A,\phi}}
\,,
\label{eq:chisa}
\end{align}
So, the two stability conditions,
Eqs.~\eqref{eq:WS2better} and~\eqref{eq:WSQbetter}, are now
\begin{align}
C_{A,\phi}> 0
\,,\label{eq:cphia}
\end{align}
\begin{align}
\chi_{S,A} C_{A,\phi}>0\,,\label{eq:cond2}
\end{align}
respectively.
This analysis to obtain the stability conditions is equivalent 
to the requirement that the matrix of variances in the grand 
canonical ensemble is positive definite. This matrix contains the 
variances $\Delta E^2$, $\Delta Q^2$ and the correlation 
$\Delta E \Delta Q$, where $E$ and $Q$ are the quantities 
that are exchanged with the heat reservoir. 
By working out the conditions of positive
definiteness, one recovers also the conditions Eqs.~\eqref{eq:cphia}
and~\eqref{eq:cond2}.

For the electrically charged black hole
in the cavity, one can compute
Eq.~\eqref{eq:chisa}
to find
$\chi_{S,A} = \frac{(d-3)\Omega r_+^{d-3}
  (1 - \frac{r_+^{d-3}}{R^{d-3}})}
  {(1 - (1 -
  \Phi^2)(\frac{r_+}{R})^{d-3})^{\frac{3}{2}}}$.
This adiabatic susceptibility is positive 
for all physical 
configurations of the charged black hole. 
Therefore, the two conditions for stability
are reduced to a single one
given in Eq.~\eqref{eq:cphia},
$C_{A,\phi}>0$.
Now,
$C_{A,\phi} = \frac{\Omega (d-2) T
r_+^{d-3}}{4}\frac{\partial r_+}{\partial T}$,
where 
$\frac{\partial r_+}{\partial T}$
can be computed, such that 
Eq.~\eqref{eq:cphia} yields
\begin{align}
  C_{A,\phi} = \frac{A (d-3)^2(d-2) x^{d-4} (1 -
  \Phi^2)^2}{32 (\pi R T)^2 ((d-1)(1 - \Phi^2)x^{d-3} - 2)}> 0\,\,,
  \label{eq:heatcapacityPhibetter}
\end{align}
with the 
dependence on the variable $x = \frac{r_+}{R}$ 
being  maintained for readability.
With
Eq.~\eqref{eq:heatcapacityPhibetter}
we recover
Eq.~\eqref{eq:stabilitysimple} for thermodynamic stability.
See also Appendix~\ref{canthermo}
for further  thermodynamic
relations. We note that for the 
case of $\Phi^2 = 0$, $C_{A,\phi}$ becomes the heat capacity 
at constant area $C_{A}$ with the expression given in
\cite{Andre:2021}. In this uncharged case, 
the bifurcation and marginal stability
radius and the photon sphere radius are the same.
A comparison between the  bifurcation and marginal stability
radius and the photon sphere radius
is done in Appendix~\ref{photonandbhcdahlandreasson},
showing that these radii do not coincide. This means
that 
the connection displayed in the
uncharged case
is not generic, it holds only in the pure
gravitational situation.

A comparison with the self-gravitating static electrically charged
thin shell in $d$-dimensions studied in \cite{Fernandes:2022} is worth
doing. Indeed, it is remarkable that the thermodynamic pressure given
in Eq.~\eqref{eq:meanpressure} and the thermodynamic energy given in
Eq.~\eqref{eq:Energy} in the grand canonical ensemble have the same
expression as the matter pressure and the matter rest mass, or rest
energy, of the corresponding self-gravitating charged spherical shell
in equilibrium. Additionally, by choosing for the matter of the thin
shell the equations of state corresponding to the temperature and
electric potential of the black hole, the shell will also have the
Bekenstein-Hawking entropy and its stability at constant area is given
by the same condition, i.e., positive heat capacity at constant electric
potential.

\subsection{Most favorable thermodynamic phase and
phase transitions}

In a thermodynamic system characterized by the grand canonical
potential $W$, it happens that all spontaneous processes always occur
in the sense of decreasing $W$.  The configuration we are studying is
a black hole inside a reservoir characterized by a fixed area $A$, a
fixed temperature $T$, and a fixed electric potential $\phi$.  So,
thermodynamically $W$ is the most suited thermodynamic potential to use
in this problem.  One feature here is that, there are no restrictions
on the energy $E$, and on the electric charge $Q$, which can flow
through the boundary with area $A$.  Thus, it is relevant to know
whether the stable black hole is the thermodynamic state with less
energy $W$, or there is another state to which the black hole can make
a phase transition.  One sees that now one uses the thermodynamic
language, and so uses phase transitions instead of quantum transitions
as we did previously.  But the results are the same, as here one uses
$W$ instead of $I_0$, with $TI_0=W$.  We summarize the results using
the grand canonical potential $W$.

In the uncharged case, one has $W_{\rm hot\, flat\, space} = 0$ and so
the black hole is favored or not depending on whether
the black hole with horizon radius $r_{+2}$ has
a $W$ lower or greater than zero.
It has been found that the radius where $r_{+2}$
yields $W=0$ is equal to the Buchdahl radius, $r_{\rm Buch}$.
For radii $r_{+2}$ higher than $r_{\rm Buch}$ the black hole
is favored. 
In the electrically charged case, one has $W_{\rm hot\, flat\, space}
= 0$, and corresponds to a cavity without a black hole and
without charge. One can emulate  hot flat space by a
tiny electric hot sphere in flat space
to find $W_{\rm hot\, sphere}$,
tends to zero as the radius of the sphere tends to zero.
So, essentially, in this setting
the black hole is favored when its $W$ is less than zero.
It is found that the radius where $r_{+2}$
yields $W=0$ is not related to the 
Buchdahl-Andr\'easson-Wright radius, a generalization for the
Buchdahl radius to any higher dimension $d$
that includes electric charge, see
Appendix~\ref{photonandbhcdahlandreasson}.
There is also the extreme black hole solution  localized at the radius of the
cavity. It is found that the stable black hole $r_{+2}$ has
always lower or equal $W$ than 
$W_{\rm
extreme\, black\, hole}$, and therefore the stable black hole is always
more favorable than the extremal black hole with horizon at the
cavity.

\section{The case $d=5$: Zero loop approximation
and thermodynamics}
\label{d=5}

\subsection{$d=5$ in the zero loop approximation}

\subsubsection{Reduced action}

Here we apply the whole formalism to $d=5$ dimensions. The $d=4$ case
is done in the Appendix~\ref{app:d4case} where a comparison with
\cite{Braden:1990} is carried out.  In $d=5$, the reduced action taken
from Eq.~\eqref{eq:reducedaction} is
\begin{align}
I_* = \frac{3\pi}{4}\beta R^{2}\left(1-\sqrt{f}\right) -
q \beta \phi - \frac{\pi^2 r_+^{3}}{2}\,,
\label{eq:reducedaction5d}
\end{align}
where 
\begin{align}
f =
\left(1-\frac{r_+^{2}}{R^{2}}\right)\left(1-
\frac{1}{3\pi^3}
\frac{
q^2}{r_+^{2} R^{2}}\right).
\label{f5d}
\end{align}
with $I_*=I_*[\beta,\phi, R; r_+,q]$
and $f=f[R;r_+,q]$. For $d=5$,
one has $\Omega=2\pi^2$ and, since
$\lambda = \frac{8\pi}{(d-2)(d-3)\Omega^2}$, one has 
$\lambda = 
\frac{1}{3\pi^3}$, which are quantities that
have been used and will be used along this section.

\subsubsection{Stationary points and
analysis of the stationary points}

The horizon radius solutions
$r_+$ obey a relation given by
Eq.~\eqref{eq:extremaaction}
which for $d=5$ becomes
\begin{align}
(1-\Phi^2)\left(\frac{r_+}{R}\right)^4 -
\left(\frac{r_+}{R}\right)^2 + (1 - \Phi^2)^2\frac{1}{(2\pi
RT)^2}=0\,,\label{eq:extremaaction5d}
\end{align}
where
here $\gamma=(2\pi RT)^2\frac{\Phi^2}{(1-\Phi^2)^2}$.
The electric charge $q^2$ obeys a relation given by
Eq.~\eqref{eq:extrema2action} which now turns into
\begin{align}
\frac{\abs{q}}{R^2}=2\sqrt{3}\,\pi^{\frac{5}{2}}\, 
\frac{RT\abs{\Phi}}{1-\Phi^2}\left(\frac{r_+}{R}\right)^3\,.
\label{eq:extrema2action5d}
\end{align}
Clearly, the stationary points, i.e., the solutions
for the horizon radius $r_+$ and the electric charge $q$
are functions of $T$, $\Phi$, and $R$, i.e.,
one has $r_+=r_+(T,\Phi,R)$ and $q=q(T,\Phi,R)$.

\vfill
\vskip 10cm

Since Eq.~\eqref{eq:extremaaction5d} can be reduced
to a quadratic equation, a property of
$d=5$, one can obtain
from Eqs.~\eqref{eq:extremaaction5d}
and~\eqref{eq:extrema2action5d}
the analytic expressions for the $r_+$ and $q$
of the two stationary points that
yield the
two
electrically charged black hole solutions.
The first stationary point or solution, $r_{+1}$, is the small
black hole
and is given by
\begin{align}
\frac{r_{+1}}{R} = \frac{1}{\sqrt{2(1-\Phi^2)}}
\left[1 - \sqrt{1 -
\frac{(1 - \Phi^2)^3}{(\pi R T)^2}}
\right]^{\frac{1}{2}}\,,
\label{solutionr+1}
\end{align}
\begin{align}
\frac{\abs{q_1}}{R^2}=
\sqrt{\frac32}\, 
\frac{\pi^{\frac{5}{2}} RT\Phi}{(1-\Phi^2)^{\frac52}}
\left[1 - \sqrt{1 -
\frac{(1 - \Phi^2)^3}{(\pi R T)^2}} \right]^{\frac{3}{2}}\,,
\label{solutionq1}
\end{align}
This solution $r_{+1}$ given in Eq.~\eqref{solutionr+1} with negative
sign in the square root was designated $x_{1}$ in
Sec.~\ref{sec:Stationarypointsanalysis}. Here,
we keep the $r_{+1}$ notation.
The second stationary point
or solution, $r_{+2}$, is the large
black hole and is given by
\begin{align}
\frac{r_{+2}}{R} = \frac{1}{\sqrt{2(1-\Phi^2)}}\left[1 + \sqrt{1 -
\frac{(1 - \Phi^2)^3}{(\pi R T)^2}} \right]^{\frac{1}{2}}\,,
\label{solutionr+2}
\end{align}
\begin{align}
\frac{\abs{q_2}}{R^2}=
\sqrt{\frac32}\, 
\frac{\pi^{\frac{5}{2}} RT\Phi}{(1-\Phi^2)^{\frac52}}
\left[1 + \sqrt{1 -
\frac{(1 - \Phi^2)^3}{(\pi R T)^2}} \right]^{\frac{3}{2}}\,.
\label{solutionq2}
\end{align}
The solution $r_{+2}$ given in Eq.~\eqref{solutionr+2} with positive
sign in the square root was designated by $x_2$ in
Sec.~\ref{sec:Stationarypointsanalysis}.  Here, we keep the
$r_{+2}$ notation.  The condition for the two black hole solutions to
exist is given by Eq.~\eqref{twosolutionscond}, which
reduces to
\begin{align}
0\leq (1 - \Phi^2)^3\leq (\pi R T)^2
<\infty\,,
\label{range}
\end{align}
in $d=5$ dimensions.
For zero electric charge and so $\Phi=0$, i.e., the
uncharged case, Eq.~\eqref{range}
turns into $1\leq (\pi R T)^2 <\infty$ of the
$d=5$ Schwarzschild-Tangherlini black hole, see
\cite{Andre:2020}.

Before embarking in a careful analysis of the stationary points,
it is useful to
make an analysis of the limits.
First, 
for very large $\pi RT$, $(\pi R T)^2\to\infty$,
independently of $\Phi$, 
the solution $r_{+1}$ behaves as $\frac{r_{+1}}{R} \rightarrow
\frac{(1 - \Phi^2)}{2\pi R T}$,
and since $\abs{\Phi}< 1$ the solution always exists.
For very large $\pi RT$, $(\pi R T)^2\to\infty$,
independently of $\Phi$, 
the solution $r_{+2}$
behaves as $\frac{r_{+2}}{R}
\rightarrow \frac{1}{\sqrt{1 - \Phi^2}}$, which for values of $\Phi^2
< 1$, as it is always the case, one has $r_{+2}> R$, so
the solution is
unphysical. This situation is different from the uncharged case,
where the solution with larger mass, $r_{+2}$, only meets the cavity at
infinite temperature, while in the charged case, the solution $r_{+2}$
meets the cavity at finite temperature, as we
are seeing here.
Second,
for $\Phi^2 \rightarrow 1$,
independently of $(\pi R T)^2$,
the
solution $r_{+1}$ tends to $r_{+1}\rightarrow 0$.
For $\Phi^2 \rightarrow 1$,
independently of $(\pi R T)^2$,
the solution $r_{+2}$ tends to
$r_{+2}\rightarrow \infty$, and so is unphysical.

\begin{figure}[t]
\centering
\includegraphics[scale=0.40]{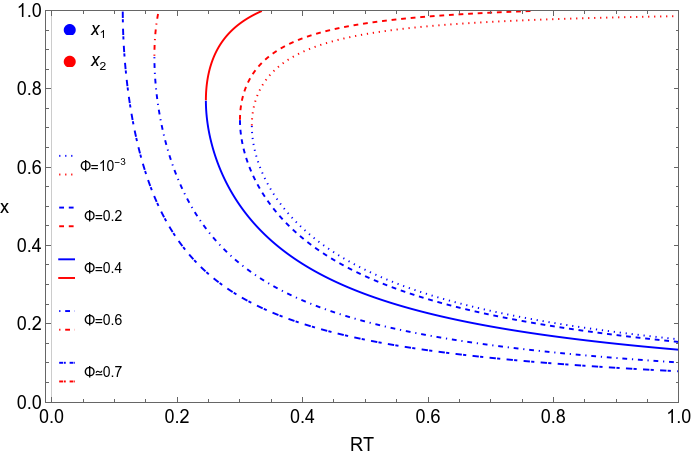}
\vskip 0.8cm
\includegraphics[scale=0.40]{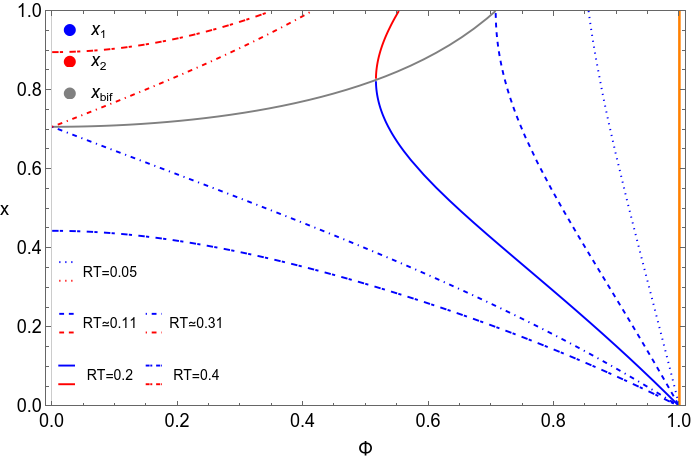}
\caption{Top plot:
Stationary points $\frac{r_{+1}}{R}=x_1$ (in blue) and
$\frac{r_{+2}}{R}=x_2$ (in red) of the reduced action $I_*$ as a
function of $RT$, for dimension $d=5$, and for five values of $\Phi$,
namely, $\Phi = 0.001$ in dotted lines, $\Phi = 0.2$ in dashed lines,
$\Phi = 0.4$ in solid lines, $\Phi = 0.6$ in dot dashed lines and
$\Phi = \frac1{\sqrt{2}}=0.7$, the last equality
is approximate, in dot
double dashed lines.
Bottom plot: Stationary points $\frac{r_{+1}}{R}=x_1$ (in blue) and
$\frac{r_{+2}}{R}=x_2$ (in red) of the reduced action $I_*$ as a
function of $\Phi$, for dimension $d=5$, and for five values of $RT$,
namely, $RT = 0.05$ in dotted lines, $RT = \frac{1}{2\sqrt{2}\pi} =
0.112$, the last equality is approximate, in dashed lines, $RT = 0.2$
in solid lines, $RT = \frac1\pi= 0.318$, the last equality is
approximate, in dot dashed lines, and $RT=0.4$ in dot double dashed
lines. The gray line corresponds to the points $x$ and $\Phi$ where
the solutions $x_1$ and $x_2$ coincide.  The orange line corresponds
to $\Phi = 1$, which is the maximum possible electric potential.
See text for further details.}
\label{fig:2solsd=5phi5piRT}
\end{figure}

\begin{figure}[t]
\centering
\includegraphics[scale=0.60]{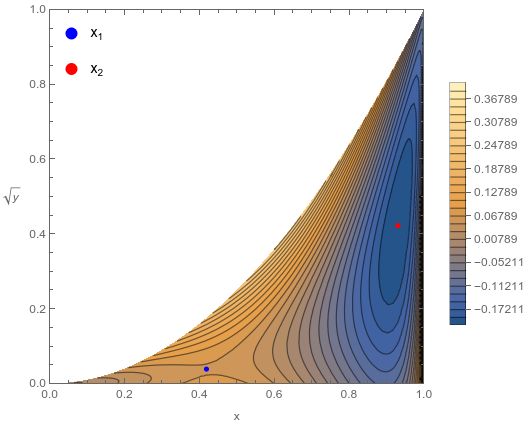}
\caption{Contour plot of the reduced action in $d=5$ dimensions,
specifically of $\frac{4 I_*}{3\pi R^{2}}$, in function of
$\frac{r_{+}}{R}=x$ and $\frac{\abs{q}}{\sqrt{3\pi^3}R^2}=\sqrt{y}$,
for $\Phi = 0.2$ and $RT = 0.4$. The blue dot is a saddle point and
corresponds to $\frac{r_{+1}}{R}=x_1$, and the red dot is a minimum
and corresponds to $\frac{r_{+2}}{R}=x_2$.
See text for further details.}
\label{fig:ContourPlotRedact}
\end{figure}

\begin{figure}[t]
\centering
\includegraphics[scale=0.33]{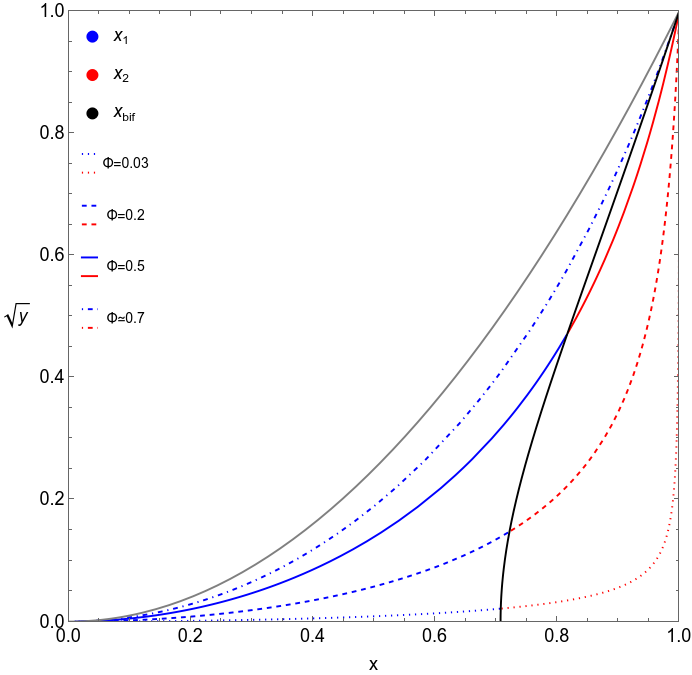}
\vskip 0.8cm
\includegraphics[scale=0.33]{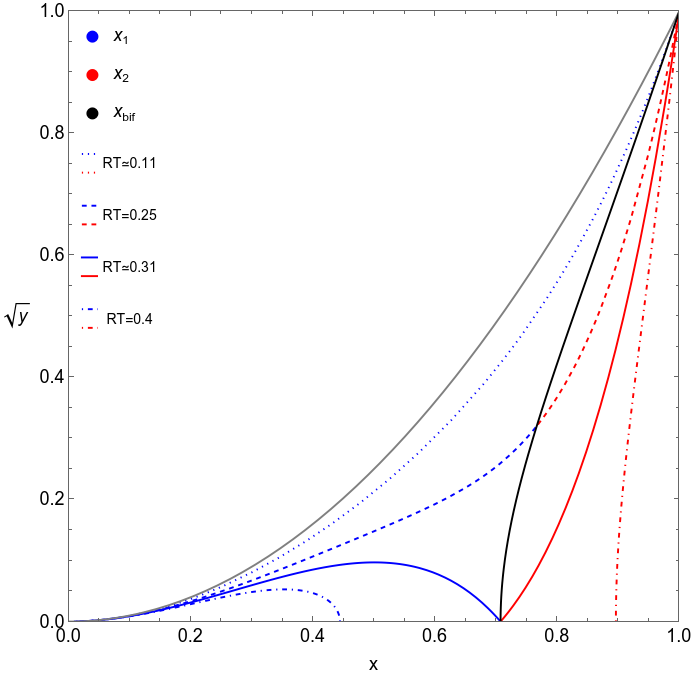}
\caption{
Top part: Path of the solutions $\frac{r_{+1}}{R}=x_1$ (in blue) and
$\frac{r_{+2}}{R}=x_2$ (in red) in the $x\times \sqrt{y}$ plane, for
$d=5$, with $x$ and $\sqrt{y}$ parametrized by $RT$, for $\Phi= 0.03$
in dotted lines, $\Phi = 0.2$ in dashed lines, $\Phi = 0.5$ in solid
lines and $\Phi = \frac1{\sqrt{2}}= 0.707$, the last equality being
approximate, in dot dashed lines.  The gray line corresponds to
extremal black holes, i.e., $\sqrt{y} = x^2$.  The black line
corresponds to the points $x$ and $\sqrt{y}$ where $x_1$ and $x_2$
coincide.  Bottom part: Path of solutions $x_1$ (in blue) and $x_2$
(in red) in the $x\times\sqrt{y}$ plane, for $d=5$, with $x$ and
$\sqrt{y}$ parametrized by $\Phi$, for
$RT=\frac{1}{2\sqrt{2}\pi}=0.11$, the last equality being approximate,
in dotted line, $RT = 0.25$ in dashed lines, $RT = \frac1\pi=0.3$, the
last equality being approximate, in solid lines, and $RT = 0.4$ in dot
dashed lines.  The gray line corresponds to the condition of extremal
black holes, i.e., $\sqrt{y} = x^2$.  The black line corresponds to
the points $x$ and $\sqrt{y}$ where $x_1$ and $x_2$ coincide.
See text.
}
\label{fig:ContourPlotRedactPathRTPhi}
\end{figure}

A careful analysis of the stationary points presented in solution
$r_{+1}$ of Eqs.~\eqref{solutionr+1}-\eqref{solutionq1} and in
solution $r_{+2}$ of Eqs.~\eqref{solutionr+2}-\eqref{solutionq2} is
now given.  Several plots are made in
Figs.~\ref{fig:2solsd=5phi5piRT}$-$\ref{fig:ContourPlotRedactPathRTPhi}
that complement Eqs.~\eqref{solutionr+1}-\eqref{range} and the main
text.
Note that $\frac{d-3}{d-1}=\frac12$ in $d=5$, and so
from Eq.~\eqref{stronger} one has that
the value $\Phi^2=\frac12$ plays an important role in the
analysis. 
Thus, we divide the analysis into two parts,
namely, $0\leq\Phi^2 \leq \frac{1}{2}$
and $\frac{1}{2}<\Phi^2 <1$.

\vskip 0.3cm
\noindent
(i) For $0\leq\Phi^2 \leq \frac{1}{2}$, 
there are three branches.

\noindent
(a) For $0\leq (\pi R T)^2 <(1-\Phi^2)^3$, there are no
stationary points, and so no black hole
solutions, only hot flat space, see below.

\noindent
(b)
For $(1-\Phi^2)^3 \leq (\pi R T)^2 \leq \frac{(1 -
\Phi^2)^2}{4\Phi^2}$,
both black hole solutions lie inside
the cavity, i.e., $r_{+1}\leq R$ and $r_{+2}\leq R$.
In the case of the equality on the right side, i.e.,
$(\pi R T)^2 = \frac{(1 - \Phi^2)^2}{4\Phi^2}$, the
solution $r_{+1}$ obeys $r_{+1}<R$, and the solution $r_{+2}$
satisfies $r_{+2}=R$ with the charge $q_2$ obeying $\abs{q_2}=
\sqrt{3 \pi^3}\,r_+^2$, i.e., it is maximal, which means that the $r_{+2}$
solution is an extremal electrically charged black hole.
The particular case $\Phi^2 = \frac{1}{2}$
yields that $(1-\Phi^2)^3=\frac{(1 -
\Phi^2)^2}{4\Phi^2}=\frac18$, so that 
$(\pi R T)^2 =\frac18$, and now the
$r_{+1}$ and $r_{+2}$ solutions merge into one,
an
extremal electrically charged black hole that
obeys $r_{+1}=r_{+2} = R$.

\noindent
(c)
For $\frac{(1 - \Phi^2)^2}{4\Phi^2}<
(\pi R T)^2 <\infty$,
 the solution $r_{+1}$
has always $r_+<R$ and so exists. For $\Phi$ near zero,
$r_{+1}$ is small and as the
value of $(\pi R T)^2$
increases  $r_{+1}$ goes to zero.
For $\Phi$ near $\frac12$ from below, 
$r_{+1}$ is near $R$ 
and as  $(\pi R T)^2$
increases $r_{+1}$ goes to zero.
On the other hand, the solution $r_{+2}$
obeys $r_{+2}> R$,
so is unphysical.
This situation is different from the uncharged case, where the
solution with larger mass, $r_{+2}$, only meets the cavity at infinite
temperature, while in the charged case, the solution $r_{+2}$ meets the
cavity at finite temperature.

\vskip 0.3cm
\noindent
(ii) For $\frac{1}{2}< \Phi^2 < 1$, there are three branches.

\noindent
(a) For $0\leq (\pi R T)^2 <(1-\Phi^2)^3$, there are no black hole
solutions, only hot flat space, see below.

\noindent
(b)
For $(1-\Phi^2)^3 \leq (\pi R T)^2 \leq \frac{(1 -
\Phi^2)^2}{4\Phi^2}$,
 both solutions $r_{+1}$ and $r_{+2}$
lie outside the cavity and so are unphysical.
This means that within this range there are no black hole
solutions, presumably only hot flat space, see below.

\noindent
(c) For $\frac{(1 - \Phi^2)^2}{4\Phi^2}< (\pi R T)^2 <\infty$, the
solution $r_{+1}$ starts at $r_{+1} = R$ in the case of $(\pi R T)^2 =
\frac{(1 - \Phi^2)^2}{4\Phi^2}$ and then decreases toward zero as the
temperature increases.  On the other hand, the solution $r_{+2}$
remains outside the cavity and so is unphysical.

In Fig.~\ref{fig:2solsd=5phi5piRT} top, the plots of the two
stationary points $\frac{r_{+1}}{R}$ and $\frac{r_{+2}}{R}$ given in
Eqs.~\eqref{solutionr+1} and \eqref{solutionr+2} as functions of $RT$
for five values of $\Phi$ are shown.  In
Fig.~\ref{fig:2solsd=5phi5piRT} bottom, the plots of the two
stationary points $\frac{r_{+1}}{R}$ and $\frac{r_{+2}}{R}$ given in
Eqs.~\eqref{solutionr+1} and \eqref{solutionr+2} as functions of
$\Phi$ for five values of $RT$ are shown.  In these plots,
the quantity $\Phi$ was chosen instead of $\phi$ so that the
comparison between the analytical study and the plots is
straightforward.  $\Phi$ is proportional to
$\phi$, specifically
in $d=5$ one has $\Phi = \sqrt{\frac{16\pi}{3}} \phi$, then
$\Phi$ is fixed as $\phi$ is fixed.

\vfill
\vskip 1cm

In Fig.~\ref{fig:ContourPlotRedact}, a contour plot of the reduced
action $I_*$,
given in Eq.~\eqref{eq:reducedaction5d},
for $RT = 0.5$ and $\Phi = 0.2$, as a function
of $\frac{r_{+}}{R}=x$ and
$\frac{\abs{q}}{\sqrt{3\pi^3}R^2}=\sqrt{y}$, displays the two stationary
points $\frac{r_{+1}}{R}=x_1$ and $\frac{r_{+2}}{R}=x_2$ in this case.
One can see the nature of the stationary points in the contour plot, 
with $r_{+1}$ being a saddle point and $r_{+2}$ being a minimum, this 
is proven generally in the next section. Although the effect on the 
contour plot for other values of $RT$ and $\Phi$ is not plotted here, 
we present in the
next plot the effect on the migration of the stationary points.

In Fig.~\ref{fig:ContourPlotRedactPathRTPhi} top, the migration path
of the two stationary points $\frac{r_{+1}}{R}=x_1$ and
$\frac{r_{+2}}{R}=x_2$ from a point in the central region where they
coincide to the two points at the corners is shown as a function of
$RT$ for four different values of $\Phi$.  The
gray line corresponds to the condition of extremal black holes,
namely,
$\sqrt{y} = x^2$, i.e., $\frac{|q|}{\sqrt{3\pi^3}} = r_+^2$.  The black
line corresponds to the points $x$ and $\sqrt{y}$ where the solutions
$x_1$ and $x_2$ coincide. For the minimum possible temperature in each
case, the solutions are at the black line, and as one increases the
temperature, $x_1$ decreases toward the origin $x=\sqrt{y}=0$, where
$RT \rightarrow +\infty$, and $x_2$ increases toward $x=\sqrt{y}=1$,
where $RT \rightarrow \frac{(1-\Phi^2)^2}{4\Phi^2}$.
In Fig.~\ref{fig:ContourPlotRedactPathRTPhi} bottom, the migration
path of the two stationary points $\frac{r_{+1}}{R}=x_1$ and
$\frac{r_{+2}}{R}=x_2$ from a point in the central region where they
coincide to the two points at the corners is shown as a function of
$\Phi$ for four different values of $RT$.  In these plots, the
quantity $\Phi$ was chosen instead of $\phi$ so that the comparison
between the analytical study and the plots is straightforward.
Since $\Phi =  \sqrt{\frac{16\pi}{3}} \phi$, one has that $\Phi$ is
fixed as $\phi$ is fixed.
The gray line corresponds to the condition of
extremal black holes, namely $\sqrt{y} = x^2$, i.e.,
$\frac{|q|}{\sqrt{3\pi^3}} = r_+^2$.  The black line corresponds to the
points $x$ and $\sqrt{y}$ where solutions $x_1$ and $x_2$
coincide. For minimum potential, the solutions either start from the
black line where the solutions coincide or start separated in
$\sqrt{y}=0$ line. As one increases the potential, $x_1$ tends to the
origin $x=\sqrt{y}=0$, where $\Phi \rightarrow 1$, and $x_2$ tends to
$x=\sqrt{y}=1$, where $\Phi \rightarrow \sqrt{(\pi RT)^2 +1}-\pi RT$.

\subsubsection{Perturbations around the zero loop
approximation and stability analysis}

Using Eq.~\eqref{eq:stabilitysimple}
with $x\equiv\frac{r_+}{R}$, one finds that
the solutions are stable when 
\begin{align}
\frac{\left(4(1 - \Phi^2)\left(\frac{r_+}{R}\right)^{2} - 2\right)
\left(1 - \left(\frac{r_+}{R}\right)^2\right)}
{(1 - (1 - \Phi^2) \left(\frac{r_+}{R}\right)^2)}>0\,\,,
\label{eq:stabilitysimple5d}
\end{align}
for $d=5$.
The physical range is $\frac{r_+}{R}<1$.
Therefore, the solutions are stable if 
$r_+ >  r_{+ \rm bif}$,
where $r_{+\rm bif}
=\frac{R}{\sqrt{2(1-\Phi^2)}}$ is the
bifurcation radius from which the solutions $r_{+2}$ and
$r_{+1}$ bifurcate at 
$(\pi RT)^2 = (1 - \Phi^2)^3$. So, one has always that
$r_{+1}< r_{+ \rm bif}< r_{+2}$.

We can spell the consequences of Eq.~\eqref{eq:stabilitysimple5d}
in more detail.
For $r_{+1}$, this means that for $(1-\Phi^2)^3 \leq (\pi R T)^2 \leq
\frac{(1 - \Phi^2)^2}{4\Phi^2}$, in the case $0\leq \Phi^2 \leq
\frac{1}{2}$, the solution does not obey the stability condition, and
so is thermodynamically unstable, and in the case $\frac12< \Phi^2 <1$
the solution $r_{+1}$ does not physically exist as it lies outside the
cavity.  For $\frac{(1 -\Phi^2)^2}{4\Phi^2}< (\pi R T)^2 <\infty$ and
$0\leq \Phi^2 <1$, the solution $r_{+1}$ does not obey the stability
condition, and so is thermodynamically unstable. Moreover, $r_{+1}$
corresponds to a saddle point of the action.
For $r_{+2}$, this means that for $(1-\Phi^2)^3 \leq (\pi R T)^2 \leq
\frac{(1 - \Phi^2)^2}{4\Phi^2}$, in the case $0\leq \Phi^2 \leq
\frac{1}{2}$, the solution obeys the stability condition, therefore
for this range of parameters the solution is thermodynamically stable,
and it is also  a
minimum of the action.
In the case $\frac12< \Phi^2 <1$ the solution  $r_{+2}$
does not physically
exist, as it lies outside the cavity.  For $\frac{(1 -
\Phi^2)^2}{4\Phi^2}< (\pi R T)^2 <\infty$ and $0< \Phi^2 < 1$ the
solution $r_{+2}$ does not physically exist also, being located
outside the cavity.

\subsubsection{Most probable configurations}
\label{mostprobable}

We now want to make a study of the most probable configurations in the
case $d=5$.  For that, we deal with stable solutions only.  The system
with less $I_0$, which is found from $I_*$ in
Eq.~\eqref{eq:reducedaction5d} and its stationary points, is the most
probable system since it gives the most important contribution to the
partition function $Z$.

To search for the most probable configuration, one has
essentially to
make a comparison of the
stable black hole with a charged equivalent hot flat space.
The action has two stable stationary points, namely,
the stationary point
$r_{+2}$ related to the stable black hole,
and the stationary point $r_+ = 0$
and $q = 0$, which corresponds to a cavity without a black hole and
without charge. The action also has 
a critical point, $r_+=R$ and
${\frac{q}{\sqrt{3\pi^3}}} =R$, so that $r_+=
{\frac{q}{\sqrt{3\pi^3}}} =R$, which corresponds to
an extremal black hole with the horizon localized at the radius of the
cavity.

To have a model for the stationary point $r_+ = 0$ and $q = 0$, we have
proposed that a nongravitating perfect conductor hot sphere with radius
$r_{\rm hs}$, inside the reservoir at constant $\beta$ and $\phi$, will do.
The electric potential is now $\phi =
\frac{q}{4\pi^2}\left(\frac{1}{r_{\rm hs}^2}- \frac{1}{R^2}\right)$, see also
Eq.~\eqref{eq:phiexpress}.  So that the action for a hot sphere,
as a model of hot flat space, in five dimensions is
\begin{align}
I_{\rm hot \,sphere} = - \frac12\, \frac{4\pi^2}
{\frac{1}{r_{\rm hs}^2}-
\frac{1}{R^2}}\beta\phi^2
\,.
\label{Ihotflatsphere5d}
\end{align}
One can then compare the action of the conducting hot sphere given in
Eq.~\eqref{Ihotflatsphere5d} with the action of the stable
configuration of the charged black hole given in 
Eq.~\eqref{eq:reducedaction5d} together
with Eqs.~\eqref{solutionr+2} and
\eqref{solutionq2}.  Clearly, from
Eq.~\eqref{Ihotflatsphere5d} one sees that for small $r_{\rm hs}$, which is
the case analogous to hot flat space, then $I_{\rm hot\,sphere}
=0$, or approximately zero, so that essentially $I_{\rm hot\,sphere}
=I_{\rm hot \,flat\,space}$ in this case.

Now, the stable black hole has positive action
only in a small range of low temperatures, namely,
for temperatures near the minimum
temperature for which the stable black hole exists.
For higher temperatures, the action for the stable black hole
is negative.
Therefore, one finds
that the small charged sphere that emulates hot flat space is less
probable for a large interval of temperatures when compared with the
stable black hole. In fact, when the solution of the stable black hole
obeys
\begin{align}
\frac{r_{+2}}{R}\geq \mu m + \sqrt{\mu^2 m^2 -
\frac{q^2}{3\pi^3}
}\,,
\label{bhmoreprobable5d}
\end{align}
with $\mu=\frac{4}{3\pi}$,
see Eq.~\eqref{eq:mu},
and with $\mu m = -\frac{9}{16} + \frac{15}{16}\sqrt{1 +
\frac{16}{27\pi^3}\frac{q^2}{R^4}}$
the corresponding action is negative
and the black hole is more probable than the tiny
charged sphere. Note that this radius $r_{+2}$ does not have
a connection to the Buchdahl-Andr\'easson-Wright bound,
in contrast to the uncharged case, 
see also Appendix \ref{photonandbhcdahlandreasson}
for more details.

\begin{figure}[b]
\centering
\includegraphics[scale = 0.36]{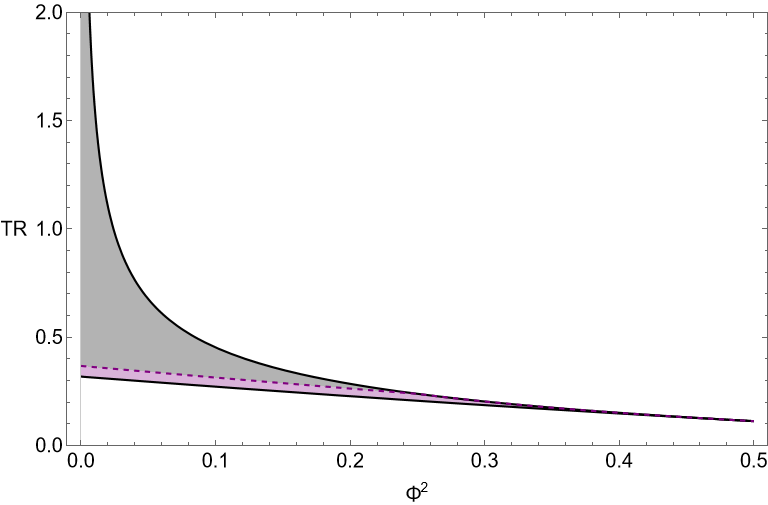}
\vskip 0.8cm
\centering
\includegraphics[scale = 0.36]{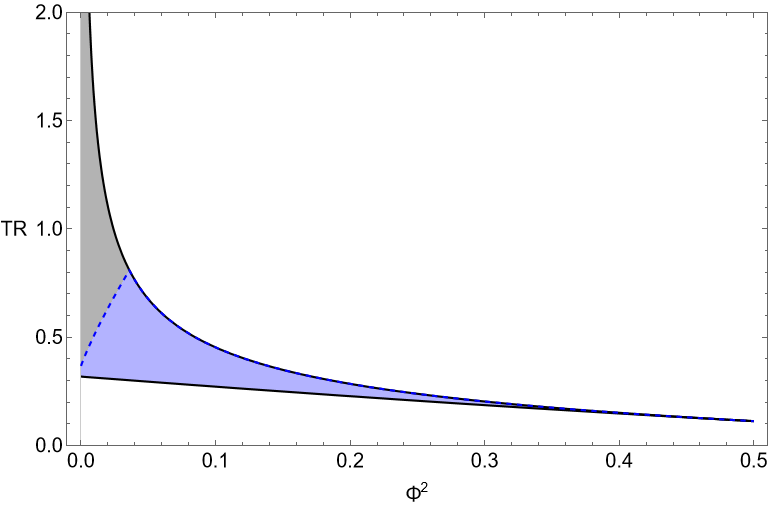}
\caption{ 
Regions of more probability in five dimensions, $d=5$,
between the stable black hole solution and the charged conducting
sphere, in function of $RT$ and $\Phi$.  Top plot: $\frac{r_{\rm hs}}{R} \to
0$.  The region in gray represents the points where the black hole
solution is more probable or favorable. The region in purple
represents the points where the infinitesimal charged conducting
sphere, emulating
electrically charged hot flat space, is
more probable. The regions in
white do not have a stable black hole solution, so
presumably
the most probable state is hot flat space.
Bottom plot:
$\frac{r_{\rm hs}}{R} = 0.99$.  The region in gray represents the points
where the black hole solution is more probable. The region in blue
represents the points where the charged conducting sphere is more
probable, with $\frac{r_{\rm hs}}{R} = 0.99$. The regions in white do not
have a stable black hole solution, so
presumably
the most probable state is hot flat space.
}
\label{fig:condspheres0and0.99}
\end{figure}

The comparison between the hot flat sphere and the stable black hole is
displayed in Fig.~\ref{fig:condspheres0and0.99}.  In the two plots of
the figure, the gray region represents the points of the pair
$(RT,\Phi)$ in which the stable black hole solution $r_{+2}$ is
more probable. The
region in purple in the top plot of the figure,
and in blue in the bottom plot,
represents the points in which the charged conducting
sphere with radius $r_{\rm hs}$ is more
probable. The regions in white represent
points where there is no stable black hole solution, so
presumably
the most probable state is hot flat space.  The upper white
region is different from the uncharged case, see \cite{Andre:2021},
because in the uncharged case the stable black hole solution exists
for temperatures up to infinite ones, whereas in the electrically
charged case, the stable black hole solution only exists within a
range of finite temperature.
In the top plot of
Fig.~\ref{fig:condspheres0and0.99},
we can see that the lower the value of $r_{\rm hs}$,
the larger the region where the
stable black hole solution is
more probable over the conducting sphere,
until the point where one has a microscopic sphere.  In that case, the
whole of the region favors the stable black hole solution, apart from
a very small region where the microscopic sphere is
more probable, see Eq.~\eqref{bhmoreprobable5d}.  This
case of a microscopic electrically charged sphere is the case that
emulates hot flat space.
In the bottom plot of
Fig.~\ref{fig:condspheres0and0.99},
we can see that the higher the value of $r_{\rm hs}$,
the smaller the region where the
stable black hole solution is
more probable, but this case is contrived,
does not emulate hot flat space.
Moreover, it must be stated that for
relatively
small values of $r_{\rm hs}$, the region of probability for the
electrically charged shell does not change much,
as even with $r_{\rm hs}
= 0.7$, the difference to $r_{\rm hs}=0$ is small. Indeed, only
variations of $r_{\rm hs}$ close to $R$ changes substantially the region of
the most probable state.
Note also that for the uncharged case, in $d=5$, the comparison
between the stable black hole solution and hot flat space was made in
\cite{Andre:2020}.  The stable black hole and the hot flat space
solution are the stationary points of the reduced action.  The most
probable state is the one with the lowest value of the action.  In the
case of the black hole, the value of the action $I_0$ depends on
$\beta$, while in the case of hot flat space one has $I_{\rm hot\,
flat\, space} = 0$.  In \cite{Andre:2020}, it was shown that $I_0 <
I_{\rm hot\, flat\, space}$ if $\beta$ is such that $\frac{r_+}{R} >
\frac{r_{\rm Buch}}{R}$, where $r_{\rm Buch}$ is the Buchdahl radius.
Thus, in the pure gravitational case the size of the horizon radius
for the black hole to be the dominant phase coincides with
the Buchdahl radius. This agreement does not extend to
when other fields are present, since
 the horizon radius
for the electrically
charged black hole to be the dominant phase does not coincide
with the Buchdahl-Andr\'easson-Wright radius, as we have shown.

In $d=5$ one can make now a
specific  comparison of the
stable black hole $r_{+2}$ with the critical point
given by $r_+=R$ and
$\frac{q}{\sqrt{3\pi^3}}=R$, i.e., an extremal black hole with
the horizon localized at the radius of the cavity, bearing in mind
that the precise extremality and the precise location can fluctuate by
Planck order quantities.  The gradient of the action is not defined at
this critical point but it may be smoothed up by taking in
consideration higher loops in the path integral or a different theory
of gravity, see Appendix~\ref{app:gradactioncavity}.  The action for
this critical point can be taken from Eq.~\eqref{Icoldcurvedspace} in
the $d=5$ case, i.e., $I_{\rm extreme\,black\, hole} = \frac{3\pi
R^2\beta}{4}\Bigl(1- \Bigr.  \left.\sqrt{f(R,r_+,q)}\right) -q \beta
\phi - \frac{\pi^2 r_+^3}{2}$, where $f(R,r_+,q)$ is taken from
Eq.~\eqref{fd} putting $d=5$ with $r_+$ and $q$ having extremal
values, so that $r_+=R$, $q=\sqrt{3\pi^3}\,R$, and $f(R,r_+,q)=0$.
Then,
\begin{align}
I_{\rm extreme\,black\, hole} = \frac{3\pi R^2\beta}{4}
-\sqrt{3\pi^3}\,R^2 \beta \phi - \frac{\pi^2
R^3}{2}\,.
\label{Icoldcurvedspace5d}
\end{align}
So $I_{\rm extreme\,black\, hole}$ has to be analyzed for each $R$ and
$\beta$ and compared with the action for the stable black hole
$r_{+2}$. It is found that,
in every instance, the stable black hole is a more probable
configuration than the extreme black hole with horizon at the cavity.

\subsection{Thermodynamics of the $d=5$ Reissner-Nordstr\"om
space in a cavity}

\subsubsection{Relation between action and grand canonical
potential}

In any dimension $d$, in particular in $d=5$, the grand potential $W$
has the dependence $W = W[T, \phi,
A]$, where here $A$ is the surface area of the
$3$-sphere at the boundary $\mathcal{B}$.
The correspondence between thermodynamics and the action of the
system is given by Eq.~\eqref{grandpot}, i.e.,
$W[T,\phi,A(R)] = T\,I_0[T,\phi,R]$. So,
here one has
\begin{align}
W=&
\frac{3\pi}{4} R^{2}\left(1-
\sqrt{\left(1-\frac{r_+^{2}}{R^{2}}\right)\left(1-
\frac{1}{3\pi^3}
\frac{
q^2}{r_+^{2} R^{2}}\right)}\right)
\nonumber\\
&
- T\,\frac{\pi^2 r_+^{3}}{2}-
q \phi\,,
\label{W5d}
\end{align}
see Eqs.~\eqref{eq:reducedaction5d}
and \eqref{f5d}.
The grand potential still
has the expression $W = E - ST- Q \phi$,
with $dW = -S dT - Q d\phi - p dA$
and with the first law of thermodynamics
$TdS=dE-\phi dQ+pdA$ holding, see Eq.~\eqref{grandpot}.

\subsubsection{Charge, mean energy, and pressure,
first law of thermodynamics,
and Euler equation, Gibbs\hskip0.04cm-Duhem relation,
and Smarr formula}

The physical quantities of the system such as the entropy, electric
charge, surface pressure, thermodynamic energy, and area can be given
in this case.

The entropy can be directly obtained from Eq.~\eqref{eq:entropyrp} in
$d=5$
\begin{align}
S = \frac14 A_+\,,
\label{eq:entropyrp5d}
\end{align}
which is the Bekenstein-Hawking entropy of a black
hole, where indeed $S = \frac{\pi^2 r_+^3}{2}$, with $A_+=2\pi^2
r_+^3$.

The electric charge can be computed from Eq.~\eqref{eq:meanQ}. In
$d=5$, it has the same appearance as in general $d$, i.e., for given
$T$, $\phi$, and $R$, one has $Q=q$, the
electric charge of
Reissner-Nordstr\"om space.

The gravitational thermodynamic surface pressure at $R$ can be
calculated from Eq.~\eqref{eq:meanpressure} to give
\begin{align}
 p = \frac{1}{8\pi R \sqrt{f}}&\left(\left(1-
\sqrt{
\left(1-\frac{r_+^{2}}{R^{2}}\right) \left(1-\frac{
q^2}{3\pi^3r_+^{2} R^{2}}\right)}
\right)^2
\right.\nonumber
\\
&\Big.
-\frac{q^2}{3\pi^3R^{4}}\Biggr)\,,
\end{align}
where  $f$ is given
in Eq.~\eqref{f5d}.
This tangential surface pressure acts along an
area $A$ that in $d=5$ is
$A=2\pi^2 R^3$,
see 
Eq.~\eqref{areaA}.

The mean thermodynamic energy
can be taken from Eq.~\eqref{eq:Energy}
to the $d=5$ case and
is
\begin{align}
E = \frac{3\pi R^{2}}{4}
\left(1-
\sqrt{
\left(1-\frac{r_+^{2}}{R^{2}}\right) \left(1-\frac{
q^2}{3\pi^3r_+^{2} R^{2}}\right)}
\right)\,.
\label{eq:Energy5d}
\end{align}
This is the same expression as the quasilocal energy evaluated at a
spherical shell of radius $R$.

The first law of thermodynamics,
$TdS=dE+pdA-\phi dQ$ given in Eq.~\eqref{1stlawusual},
for the system in $d=5$ holds, of course.

From Eq.~\eqref{eq:Energy5d}, one can write the energy
in terms of the
entropy $S$ of Eq.~\eqref{eq:entropyrp5d},
electric charge $Q$, and
surface area of the cavity $A$, 
as
\begin{align}
E =&
\frac{3
(2\pi^2)^{\frac{1}{3}}  A^{\frac{2}{3}}  }{8\pi}\times
\nonumber\\
&\left(1-\sqrt{\left(1-\left(\frac{4
S}{A}\right)^{\frac{2}{3}}\right)\left(1-\frac{ Q^2
(2\pi^2)^{\frac{4}{3}}}{3\pi^3(4SA)^{\frac{2}{3}}}\right)}
\right)\,.
\label{eq:energyintermsof5d}
\end{align}
One can then use the Euler's homogeneous function
theorem considering that under rescaling of its arguments, the energy
as a function has the property that $E\left(\nu S,\nu A,\nu
Q^{\frac{3}{2}}\right) = \nu^{\frac{2}{3}} E\left(S,A,
Q^{\frac{3}{2}}\right)$. We thus have an integrated version of the
first law of thermodynamics given by, see
Eq.~\eqref{eq:1stlawintegrated},
\begin{align}
\frac{2}{3}E = TS - pA + \frac{2}{3}\phi
Q\,,\label{eq:1stlawintegrated5d}
\end{align}
which is the Euler equation for
the system of a $d=5$
electrically charged black hole in a
heat reservoir.
By differentiating Eq.~\eqref{eq:1stlawintegrated5d} and considering
that $dE = TdS - pdA + \phi dQ$, one obtains
\begin{align}
  TdS - pdA + 3(SdT-Adp) + 2 Qd\phi = 0\,.
\label{eq:gibbs5d}
\end{align}
which is the Gibbs\hskip0.04cm-Duhem relation for
the $d=5$
electrically charged black hole in a
heat reservoir.
Then, the Smarr formula in $d=5$ is 
\begin{align}
m = \frac32 T_{\rm H}S + \phi_{\rm H}
Q\,,
\label{eq:smarr5d}
\end{align}
see Eq.~\eqref{eq:smarr}. Again, the Smarr
formula is valid for the small black hole
solution only.

\subsubsection{Equilibrium and stability in terms of
thermodynamic variables}

The heat capacity $C_{A,\phi} = 
T\left(\frac{\partial S}{\partial T}\right)_{A,\phi}$
is given by Eq.~\eqref{eq:heatcapacityPhibetter}. 
By setting $d=5$, one gets 
\begin{align}
C_{A,\phi}=
\frac{3 \left(\frac{r_+}{R}\right)\left(1 - \Phi^2\right)^2}
{8 \pi\, T^2 \left(2(1 -
\Phi^2)\left(\frac{r_+}{R}\right)^{2} - 1\right)}
\,,
\label{heatcapacityPhiA5d}
\end{align}
where we have that
$A$ is the area of the reservoir,
and we have used
$x=\frac{r_+}{R}$. For the link between the validity of zero loop 
approximation and the heat capacity,
see Appendix \ref{canthermo}.
So $ C_{A,\phi} > 0$ means
\begin{align}
2(1 - \Phi^2) \left(\frac{r_+}{R}\right)^2 - 1 > 0\,.
\label{stabilityheatcapacityPhiA5d}
\end{align}
Therefore we recover
Eq.~\eqref{eq:stabilitysimple5d}.

\subsubsection{Most favorable thermodynamic configurations}

The most favorable thermodynamic configuration
is found from 
the state with the lowest value of the grand potential $W$.
Since $W=TI_0$, the analysis is the same if done
in $I_0$ or in $W$.
What changes is that in $I$ one talks about the
most probable state and about quantum transitions,
and when using $W$ one talks about
the most favorable state and phase transitions. 
Since we have done the analysis in detail
for $I$ in $d=5$, it is not necessary
to do in $W$, we refer to Sec.~\ref{mostprobable}.

\section{Conclusions}
\label{sec:conc}

The grand canonical ensemble of a $d$-dimensional Reissner-Nordstr\"om
space in a cavity was built using the path integral approach. The
partition function of the space in a cavity was obtained by performing
the zero loop approximation to the path integral relative to the
Euclidean action, where only the term which minimizes the action
contributes to the path integral. There are two stationary points of
the action that correspond to a black hole in equilibrium with a heat
reservoir with the temperature and the electric potential fixed at the
boundary of the cavity. The stationary point with lower horizon radius
was shown to be unstable, while the stationary point with higher
horizon radius is stable, as has been proved for arbitrary $d$. The
corresponding values of the event horizon radius depending on the
temperature and electric potential of the two stationary points cannot
be found analytically for arbitrary dimensions. However, it is
possible to find analytically the event horizon radius for $d=5$,
where one needs to solve a quadratic polynomial.

There are some features of the stationary points in the electrically
charged case that differ from the electrically uncharged case. First,
the event horizon radius corresponding to the lowest temperature
allowed does not correspond to the photon sphere, unlike the uncharged
case, thus showing that this coincidence is really restricted
to the pure gravitational case.
Second, the larger horizon radius solution reaches the radius of
the cavity at finite temperature, unlike the uncharged case. where the
horizon radius only reaches the cavity radius at infinite temperature.

The grand canonical ensemble of the stable stationary point can be
constructed by comparing the partition function given by the path
integral with the partition function of the grand canonical
ensemble. The grand potential can be obtained from the comparison and
the thermodynamics of the black hole corresponding to the stable
stationary point is recovered. We have shown that the entropy
corresponds to the Bekenstein-Hawking entropy, the pressure
corresponds to the pressure of a self-gravitating static electrically
charged spherical thin shell in equilibrium, and the thermodynamic
energy has the same expression as the expression for the quasilocal
energy.  The first law of thermodynamics with constant area is obeyed
at the stationary points of the action, as one would expect. The
stability of the stationary points is described by the heat capacity
at constant area and electric potential. If this heat capacity is
positive, then the stationary point is stable. This fits well with the
relationship between thermodynamic stability and the heat capacity.

Additionally, we have compared the stable black hole solution to an
electrically charged conducting hot sphere in flat space, to analyze
when one is more favorable than the other. In this case,
a configuration is more favorable than the other when its grand
potential $W$ is lower. We have shown that this depends on the value
of the temperature, of
the electric potential of the
reservoir, and of the radius of the
conducting sphere. Moreover, the smaller the radius of the conducting
sphere, the larger the region where the stable black hole is
favored.
The comparison of
the Buchdahl-Andr\'easson-Wright bound radius in $d$-dimensional
Reissner-Nordstr\"om spacetimes
with the minimum radius for which the stable black hole phase
is thermodynamically favored,
dictates that they have no relationship,
thus showing
that the connection displayed in the Schwarzschild case is not
generic, rather it is a very restricted equality holding only in
the pure gravitational situation.

\section*{Acknowledgments}
We thank financial support from Funda\c c\~ao para a Ci\^encia e
Tecnologia - FCT through the project~No.~UIDB/00099/2020 and
project~No.~UIDP/00099/2020

\appendix

\section{Connection of thermodynamic radii
to spacetime radii}
\label{photonandbhcdahlandreasson}

\subsection{Thermodynamic bifurcation radius and
the photon sphere radius}

In the case
of the grand canonical ensemble of a $d$-dimensional
Reissner-Nordstr\"om black hole in a cavity,
we have seen in Eq.~\eqref{xmin}
that the two
thermodynamic black hole solutions, represented by
$r_{+1}$ and $r_{+2}$,
bifurcate
from a horizon radius obeying
$\frac{r_+}{R}=\frac{2^{\frac1{d-3}}}{\left({(d-1)}(1 -
\Phi^2)\right)^{\frac1{d-3}}}$, or in terms of $R$,
\begin{align}
R= 
\left(\frac{(d-1)}{2}(1 - \Phi^2)\right)^{\frac1{d-3}}r_+\,.
\label{horizonbifurcradius}
\end{align}
A black hole
for which the horizon radius $r_+$
satisfies Eq.~\eqref{horizonbifurcradius}
is marginally
stable to thermodynamic perturbations, black
holes with larger radius $r_+$
are thermodynamically stable.
Thus, the bifurcation radius is also
the marginal thermodynamic stable radius.

The photon sphere radius $R$ of a $d$-dimensional
Reissner-Nordstr\"om black hole
is given by, see Appendix \ref{photonsphere},
\begin{align}
R = \left(\frac{(d-1)}{2}
\left(1 + \frac{d-3}{d-2}\Phi^2\right)
\right)^{\frac1{d-3}}r_+
\,.
\label{photonsphereradius}
\end{align}
At this radius, null geodesics can have
circular trajectories, and thus photons
can execute circular orbits.

From direct comparison
between Eqs.~\eqref{horizonbifurcradius}
and \eqref{photonsphereradius}, we see
that the two radii are distinct in any dimension
$d$, so that in the grand canonical
ensemble of the Reissner-Nordstr\"om black hole
there is no connection between them.

Of course, when there is no charge and so no $\Phi$,
the two radii coincide as Eqs.~\eqref{horizonbifurcradius}
and \eqref{photonsphereradius} both yield
$R= 
\left(\frac{d-1}{2}\right)^{\frac1{d-3}}r_+$,
and so the radius of the
cavity at which
a stable black hole appears 
corresponds to the photon sphere radius.

Thus, we have verified that the equality between the bifurcation and
marginal stability radius and the photon orbit radius
only holds in the pure gravitational case,
the equality does not extend to the grand canonical ensemble of the
$d$-dimensional Reissner-Nordstr\"om black hole.

\subsection{Most favorable configuration radius and
the Buchdahl-Andr\'easson-Wright sphere radius}
\label{bawradius}

In the case of grand canonical ensemble of a $d$-dimensional
Reissner-Nordstr\"om black hole in a cavity, the stable solution has a
negative action  $I_0$, see Eq.~\eqref{eq:actionI0full},
or equivalently, a negative
grand potential $W$, see Eq.~\eqref{eq:grandpotentialactionI0full},
if
\begin{align}
&\frac{\mu m}{R^{d-3}} \leq -\frac{4 (d-2)^2}{(d-1)^2(d-3)^2}
\nonumber \\&+ \frac{2(d-2)((d-2)^2 + 1)}{(d-1)^2(d-3)^2}\sqrt{1 
+ \frac{(d-1)^2 (d-3)^2}{4(d-2)^2}
\frac{\lambda q^2}{R^{2d-6}}}\,.
\label{eq:bhzeroaction}
\end{align}
Note that this condition for $d=4$ is given by
$\frac{ m}{R} \leq -\frac{16}{9} +
\frac{20}{9}\sqrt{1 + \frac{9}{16} \frac{
q^2}{4\pi R^{2d-6}}}$.

The  Buchdahl-Andr\'easson-Wright
bound yields the minimum radius,
below which, an electrically charged
matter distribution obeying certain
conditions, in general relativity 
coupled to Maxwell electromagnetism in $d$ dimensions,
the spacetime is singular. 
This Buchdahl-Andr\'easson-Wright radius
was obtained in~\cite{Wright:2015}.
This radius 
and can also be found from our work with thin shells~\cite{Fernandes:2022} 
by imposing that the trace of the stress-energy
tensor of the matter in the thin shell
is zero. The radius is given by
\begin{align}
&\frac{ m}{R^{d-3}} = \frac{d-2}{(d-1)^2} 
+ \frac{1}{d-1}\frac{\lambda q^2}{ R^{2d-6}}\nonumber\\
& + \frac{d-2}{(d-1)^2}\sqrt{1 + (d-1)(d-3)\frac{\lambda q^2}{R^{2d-6}}}\,.
\label{eq:buchandreassonwrightddimensions}
\end{align}
For $d=4$, this is
$
\frac{\mu m}{R} \leq \left( \frac{1}{3} + \sqrt{\frac{1}{9} +
\frac{1}{3}\frac{\lambda q^2}{R^2}}\right)^2$, i.e.,
$
\frac{m}{R} \leq
\frac{2}{9}+
\frac{1}{3}\frac{ q^2}{4\pi R^2}+
       \frac23\sqrt{\frac{1}{9} +
\frac{1}{3}\frac{q^2}{4\pi R^2}}$.

From direct comparison between Eqs.~\eqref{eq:bhzeroaction} and
\eqref{eq:buchandreassonwrightddimensions}, we see that the most
favorable configuration radius and the Buchdahl-Andr\'easson-Wright
radius are distinct in any dimension $d$, and so there is no
connection between them.

Of course, when there is no charge, $q=0$, and so no $\Phi$, $\Phi=0$,
both radii are equal to Buchdahl radius.  When $q=0$, one finds that
Eqs.~\eqref{eq:bhzeroaction} and
\eqref{eq:buchandreassonwrightddimensions} both yield $\frac{r_+}{R}
\geq \left(\frac{4(d-2)}{(d-1)^2}\right)^\frac{1}{d-3}$, which is the
Buchdahl radius.  In this case, the stable solution has a negative
free energy if the radius of the black hole is larger than the
Buchdahl radius.

Thus, we have verified that the equality between the black hole
radius for which the stable solution has a zero $W$ and the
Buchdahl-Andr\'easson-Wright radius only holds in the pure
gravitational case.

\section{Gradient of the action of the two  critical points}
\label{app:gradactioncavity}

\subsection{Gradient of the action}

The gradient of the action in Eq.~\eqref{eq:reducedaction} yields
\begin{align}
&\frac{\mu}{R^{d-2}}\frac{\partial I_*}{\partial x} = 
\frac{(d-3)\beta}{2R x\sqrt{f}}\left[x^{d-3} -
\frac{y}{x^{d-3}}\right]
-2\pi x^{d-3}\,,
\label{eqapp:graddaaction}
\\
&\frac{\mu}{R^{d-2}}\frac{\partial I_*}{\partial \sqrt{y}} = 
\frac{\beta \sqrt{y}}{R x^{d-3}\sqrt{f}}(1-x^{d-3}) -
\frac{\beta \Phi}{R}\,,
\label{eqapp:graddaaction1}
\end{align}
where $x = \frac{r_+}{R}$, $y = \frac{\lambda q^2}{R^{2d-6}}$ and
$\Phi = (d-3)\Omega \sqrt{\lambda}\phi$.

\subsection{Gradient of the action of hot flat space}

Here we want to analyze the
gradient of the action
given in Eqs.~\eqref{eqapp:graddaaction}
and \eqref{eqapp:graddaaction1}
at the critical point of hot flat space. For that,
we calculate the limit of the gradient for $x=y=0$ along the curve $y
= (\eta)^{2} x^{2d-6}$, where $\eta$ is a positive constant of the
curve. Note that one must consider $\eta < 1$ so that the curve
is inside the domain of the action. This family of curves cover
the possible directions from the point $x=y=0$ to the physical
domain of the action. The gradient is given by
\begin{align}
& \frac{\mu}{R^{d-2}} \frac{\partial I_*}{\partial x} =
\frac{(d-3)\beta 
x^{d-4}}{2 R}\left(1 - \eta^2\right)\,,\\
& \frac{\mu}{R^{d-2}}\frac{\partial I_*}{\partial \sqrt{y}} =
\frac{\beta}{R}\left(\eta - \Phi\right)\,.
\end{align}
We have left the dependence in $x^{d-4}$ since it yields different
limits for the case of $d=4$ and $d>4$.  Since the gradient depends on
$\eta$, then the gradient is undefined.  Nevertheless, we can
calculate the directional derivative along the vector $v =
\frac{1}{\sqrt{1 + (d-3)^2\eta^2 x^{2d - 8}}} (1 ,
(d-3)\eta x^{d-4})^T$, which will be given by
\begin{align}
D_v I_* =& \frac{(d-3)\beta x^{d-4}}{2R}\left(1 + \eta^2 -
2\eta\Phi \right)\,\,,
\end{align}
so for $d>4$, we have that the directional derivative
vanishes. Moreover, the directional $(d-3)$th derivative is positive 
since $1 + \eta^2 -2\eta\Phi>0$ for $\eta < 1$ and $\Phi \leq 1$, and
so this can be considered as a minimum, although formally the partial
derivative in $\sqrt{y}$ is undefined. For $d=4$ case, the directional
derivative does not vanish. Yet, since $1 + \eta^2 -2\eta\Phi>0$ for 
$\eta < 1$ and $\Phi \leq 1$, one can observe
that the directional derivative is positive in the
physical domain. Therefore, the action resembles a conical potential
well at the origin and so hot flat space can be considered as a
solution.

\subsection{Gradient of the action of the extremal black hole with 
horizon at the cavity}

The gradient of the action of the extremal black hole can be given by
using Eqs.~\eqref{eqapp:graddaaction}
and \eqref{eqapp:graddaaction1}. In order to study the gradient
in the critical point $x=1$ and $y=1$, we calculate the gradient of
the action in this limit along the curve $x^{d-3}= 1 - \epsilon$ and
$y = 1 - \eta \epsilon$, where $\eta$ is a constant of the
curve and $\epsilon$ parametrizes the curve. The limit of $\epsilon
\rightarrow 0^+$ is then performed, yielding the gradient
\begin{align}
&\frac{\mu}{R^{d-2}}\frac{\partial I_*}{\partial x} = 
\frac{(d-3)\beta}{2R\sqrt{\eta - 1}}(\eta-2) -
2\pi \,,
\label{eq:gradextreme1}\\
&\frac{\mu}{R^{d-2}}\frac{\partial I_*}{\partial \sqrt{y}} = 
\frac{\beta}{R\sqrt{\eta - 1}}-\frac{\beta \Phi}{R}\,,
\label{eq:gradextreme2}
\end{align}
where $\eta > 2$ so that the curve is done along configurations 
of subextremal black holes, coming from the condition $y < x^{2d-6}$.
Since there is a dependence on the curve we choose to perform the limit,
the gradient at the extremal point is not defined. 

It is interesting to see that for $\gamma = 1$, i.e.,
$\beta =
\frac{4\pi}{d-3} \frac{\abs{\Phi}}{1 - \Phi^2}R$, the gradient
vanishes in the limit along a curve with $\frac{1}{\eta} = 1 +
\frac{1}{\Phi^2}$.  Indeed, this set of temperatures corresponds to the
stable black hole solution hitting the extremal point $x=y=1$. Of
course, this only happens in one particular curve, the gradient is
still undefined.

It is also interesting to consider the directional derivative along these 
curves, in the direction of smaller $\epsilon$. Indeed, the direction 
can be described by the vector $v = 
\frac{1}{\sqrt{1 + (d-3)^2\eta^2/4}}(1,
\frac{\eta (d-3)}{2})^T$,
and so the directional derivative yields
\begin{align}
&\frac{\mu}{R^{d-2}} D_v I_* =
\frac{\frac{\beta (d-3)}{2R}\left(2\sqrt{\eta -1} -
\eta \Phi\right)
 - 2\pi}{\sqrt{1 +
(d-3)^2\eta^2/4}} \,.
\label{B8}
\end{align}
The directional derivative depends also on $\eta$ and it can be
positive or negative. In particular, for values of $\eta$ and $\Phi$
where $\gamma_{\rm bif}(\Phi,d)< \frac{4 (\eta -1)
\Phi^2}{(1-\Phi^2)^2}\left(1 -
\frac{\eta}{2\sqrt{\eta-1}}\Phi\right)^2$ the directional derivative
in Eq.~\eqref{B8} can be positive in a region $\gamma_{\rm
bif}(\Phi,d)<\gamma < \frac{4 (\eta -1) \Phi^2}{(1-\Phi^2)^2}\left(1 -
\frac{\eta}{2\sqrt{\eta-1}}\Phi\right)^2$, with $\gamma_{\rm bif}$
given in Eq.~\eqref{eq:solExcondition}. Therefore, the action near
this critical point does not resemble a potential well.

\section{Further relations between canonical ensemble and
thermodynamics}
\label{canthermo}

\subsection{Extrema as first law of thermodynamics}

The extrema of the reduced action and stability of the two solutions
of $r_+(\beta,\phi,R)$ and $q(\beta,\phi, R)$ was analyzed in
Sec.~\ref{sec:extremastability}. Yet, the physical interpretation of
these extrema and stability does not seem explicit.  Let us do this
interpretation with the help of the thermodynamic variables.

We rewrite the reduced action
in terms of the thermodynamic variables.
If we use 
$S$ of 
Eq.~\eqref{eq:entropyrp},
$Q$ 
of Eq.~\eqref{eq:meanQ}, and 
$E$ of Eq.~\eqref{eq:Energy},
Eq.~\eqref{eq:reducedaction} turns into
\begin{align}
I_* = \beta E - S - Q \beta \phi \,,
\label{eq:generalized}
\end{align}
where the functions $E$, $S$, and $Q$ are
in a first moment seen as function of $r_+$ and $q$,
and
noting that $r_+$ and $q$ here
are not restricted. 

The conditions for extrema of the action,
and so of equilibrium, are given in Eqs.~\eqref{eq:derI1}
and \eqref{eq:derI2}. Now, 
in terms of $S$ and $Q$ one has
$\left(\frac{\partial
I_*}{\partial r_+}\right)_{\beta,R,q}=
\frac{\partial
S}{\partial r_+}
\left(\frac{\partial
I_*}{\partial S}\right)_{\beta,\phi,A,Q}$
and 
$\left(\frac{\partial
I_*}{\partial q}\right)_{\beta,R,r_+}=
\left(\frac{\partial
I_*}{\partial Q}\right)_{\beta,\phi,A,S}$.
Since $\frac{\partial
S}{\partial r_+} >0$ and $\frac{\partial
Q}{\partial q} = 1$, Eqs.~\eqref{eq:derI1}
and \eqref{eq:derI2}, i.e.,
$\frac{\partial I_*}{\partial r_+} = 0$
and
$\frac{\partial I_*}{\partial q} = 0$,
respectively, 
can now be rewritten as
$\left(\frac{\partial
I_*}{\partial S}\right)_{\beta,\phi,A,Q}  = 0$
and
$\left(\frac{\partial I_*}{\partial Q}\right)_{\beta,\phi,A,S} =0$.
Then using Eq.~\eqref{eq:generalized}, one finds
from these two equations that
$\beta \frac{\partial
E}{\partial S} -1 = 0$
and
$\beta \left(\frac{\partial E}{\partial Q} -\phi\right) =0$,
respectively,
where 
$\frac{\partial E}{\partial S} \equiv
\left(\frac{\partial E}{\partial S}\right)_{A,Q}$
and
$\frac{\partial E}{\partial Q} \equiv
\left(\frac{\partial E}{\partial Q}\right)_{A,S}$, to simplify
the notation. 
So, with $\beta=\frac1T$,
the extrema give the result
\begin{align}
&
\frac{\partial E}{\partial S} =T \,,
\label{eq:extremaas1law2}\\
&\frac{\partial E}{\partial Q}=\phi\,.
\label{eq:extremaas1law1}
\end{align}
Moreover,
$E$ of Eq.~\eqref{eq:Energy},
in the  variables $A$
of 
Eq.~\eqref{areaA},
$S$ of 
Eq.~\eqref{eq:entropyrp}, and $Q$ 
of Eq.~\eqref{eq:meanQ},
is of the form
$E=E(S,A,Q)$. It is then useful
to define a quantity, namely,
thermodynamic pressure $p$, such that
\begin{align}
\frac{\partial E}{\partial A} =-p \,,
\label{eq:noextremaas1law2}
\end{align}
where 
$\frac{\partial E}{\partial A} \equiv
\left(\frac{\partial E}{\partial A}\right)_{S,Q}$.
Since by definition $dE=\frac{\partial E}{\partial S}dS
+\frac{\partial E}{\partial A}dA+
\frac{\partial E}{\partial Q}dQ$, we obtain
from Eqs.~\eqref{eq:extremaas1law2}-\eqref{eq:noextremaas1law2}
that 
\begin{align}
dE=TdS-pdA+\phi dQ,
\label{eq:1stlawextremes}
\end{align}
which is the first law of thermodynamics.
Therefore, the
condition of extrema, or
of equilibrium, in the reduced action,
is equivalent to impose the first law of
thermodynamics to the
thermodynamic energy $E$.

The derivation of the first law of thermodynamics here relies on the
fact that the reduced action can be written as
Eq.~\eqref{eq:generalized}.  As seen in Sec.~\ref{sec:GrandCanonical},
the thermodynamic quantities obtained through the grand potential are
indeed the same as the ones considered here, since the extrema
conditions are used.

\subsection{Perturbations around the zero loop approximation
and stability of the action as thermodynamic stability}

In the path integral, the stability for the extrema is given by the
requirement that the matrix ${I_{*0}}_{ij}$, whose components are given
in Eqs.~\eqref{eq:Irprp}-\eqref{eq:Iqq}, is positive definite.  In
thermodynamics, we have to deal with second derivatives of $W$ for
stability, Eqs.~\eqref{eq:WS2}-\eqref{eq:WSQ}. Since $\beta W=I_0$ one
should be able to confront and compare the results of the two
formalism to show their equivalence in this stability aspect.

For that we rewrite that a sufficient condition
is 
\begin{align}
{I_{*0}}_{r_+ r_+}{I_{*0}}_{qq} -
\left({{I_{*0}}_{r_+ q}}\right)^2>0\,,
\label{eq:condstab2app}
\end{align}
see Eq.~\eqref{eq:condstab2}.  So we have to find the components of
the matrix ${I_{*0}}_{ij}$ by considering $I_*$ given by
Eq.~\eqref{eq:generalized}, noting that for $I_*$, $r_+$ and $q$ are
not restricted.  In addition, since we are making the transformations
from $r_+$ to $S$ and from $q$ to $Q$, specifically, $S = \frac{\Omega
r_+^{d-2}}{4}$ and $Q=q$, see Eqs.~\eqref{eq:entropyrp} and
\eqref{eq:meanQ}, we have to relate the derivatives in $S$ to
derivatives in $r_+$ and derivatives in $Q$ to derivatives in $q$. The
action $I_* = I_*(r_+,q;\beta,\phi,R)$ is now given as $I_* =
I_*(S(r_+),Q;\beta, \phi, R)$.  It is understood that a partial
derivative of $I_*$ in order to either $S$ or $Q$ means we are working
with $I_* = I_*(S(r_+),Q;\beta, \phi, R)$.  Also, we use the
transformation formulas
$\frac{\partial P(S(r_+),q)}{\partial r_+}=
\frac{\partial S}{\partial r_+}
\frac{\partial P(S(r_+),q)}{\partial S}$ and
$\frac{\partial^2 P(S(r_+),q)}{\partial r_+^2} =
\frac{\partial}{\partial r_+}\left(\frac{\partial S}{\partial r_+}
\frac{\partial P(S(r_+),q)}{\partial S} \right) = 
\frac{\partial^2 S}{\partial r_+^2}
\frac{\partial P(S(r_+),q)}{\partial S} 
+ \left(\frac{\partial S}{\partial r_+}
\right)^2 \frac{\partial^2 P(S(r_+),q)}{\partial S^2}$, 
where $P(S(r_+),Q)$ stands for any  of the functions
of interest.

To calculate ${I_{*0}}_{r_+ r_+}$ we have to calculate first
${I_{*}}_{r_+ r_+}$ with $r_+$ and $q$ 
not restricted. So we have
${I_{*}}_{r_+ r_+} =
\frac{\partial^2 S}{\partial r_+^2}
\frac{\partial I_*}{\partial S}
+
\left(\frac{\partial S}{\partial r_+}\right)^2
\frac{\partial^2 I_*}{\partial S^2}
$. Since the extrema condition implies that $\frac{\partial
I_*}{\partial r_+} = \frac{\partial S}{\partial r_+} \frac{\partial
I_*}{\partial S} = 0$,  we have then in the new variable
$\frac{\partial I_*}{\partial S} = 0$. Putting this in the
calculation above for ${I_*}_{r_+ r_+}$ and restricting $r_+$ and $q$ to
the extrema, we have
\begin{align}
{I_{*0}}_{r_+ r_+} = \left(\frac{\partial S}{\partial r_+}\right)^2_0 
\left(\frac{\partial^2 I_*}{\partial S^2}\right)_{0}\,,
\end{align}
where the prefix $0$ means the derivative is evaluated at the extrema.
To calculate ${I_{*0}}_{r_+ q}$, we start as well with ${I_*}_{r_+
q}$. The quantity ${I_*}_{r_+ q}$ in the new variables is ${I_*}_{r_+ q} =
\frac{\partial S}{\partial r_+} \frac{\partial^2 I_*}{\partial Q
\partial S}$, where the derivatives in $I_*$ can be interchangeable
and it has been used that $S = S(r_+)$ as it does not depend on $q$ or
$Q$. Evaluating at the extrema, we have then
\begin{align}
{I_{*0}}_{r_+ q} = \left( \frac{\partial S}{\partial r_+}\right)_0
\left(\frac{\partial^2 I_*}{\partial Q \partial S}\right)_0\,.
\end{align}
To calculate ${I_{*0}}_{qq}$, it is straightforward since $q=Q$
that the correspondent quantity in new variables is
\begin{align}
{I_{*0}}_{qq} = \left(\frac{\partial^2 I_*}{\partial Q^2}\right)_0\,.
\end{align}
With all the quantities in the new variables,
the stability condition,
Eq.~\eqref{eq:condstab2app}, can be written as
\begin{align}
\left(\frac{\partial^2 I_*}{\partial S^2}\right)_0
\left(\frac{\partial^2 I_*}{\partial 
Q^2}\right)_0 - \left(\frac{\partial^2 I_*}{\partial S
\partial Q} \right)^2_0 > 0\,,
\label{eq:condstab2app22}
\end{align}
where we have divided  by
$\left(\frac{\partial S}{\partial r_+}\right)^2_0$ since
it is positive.

To turn this condition, Eq.~\eqref{eq:condstab2app22}, into a
condition on thermodynamic quantities, we recall
our previous results.
Starting by $\left(\frac{\partial^2 I_*}{\partial
S^2}\right)_0$, we have that $\left(\frac{\partial^2 I_*}{\partial
S^2}\right)_0 = \left(\frac{\partial }{\partial S}\left(\beta
\frac{\partial E}{\partial S} - 1\right)\right)_0 = \beta
\frac{\partial^2 E}{\partial S^2} = \beta \left(\frac{\partial
T}{\partial S}\right)_{A,Q} = \frac{1}{C_{A,Q}}$, where
Eq.~\eqref{eq:extremaas1law2} was used and the heat capacity $C_{A,Q}$
at constant area and charge has been defined as $C_{A,Q} =T
\left(\frac{\partial S}{\partial T}\right)_{A,Q}$, with
$\beta=\frac1T$.
For $\left(\frac{\partial^2 I_*}{\partial Q \partial S}\right)_0$, we
have that $\left(\frac{\partial^2 I_*}{\partial Q \partial S}\right)_0
= \left(\frac{\partial}{\partial Q} \left(\beta \frac{\partial
E}{\partial S} - 1\right)\right)_0 = \beta \frac{\partial^2
E}{\partial Q \partial S} = \frac1T \left(\frac{\partial T}{\partial
Q}\right)_{S,A} = -\frac{\lambda_{T,A}}{C_{A,Q}}$, where
Eq.~\eqref{eq:extremaas1law2} was used and the latent heat capacity
$\lambda_{T,A}$ at constant temperature and area has been defined as
$\lambda_{T,A} = \left(\frac{\partial S}{\partial Q} \right)_{T,A}$.
For $\left(\frac{\partial^2 I_*}{\partial Q^2}\right)_0$, we have that
$\left(\frac{\partial^2 I_*}{\partial Q^2}\right)_0 =
\left(\frac{\partial}{\partial Q}\left(\beta\frac{\partial E}{\partial
Q} - \beta\phi \right) \right)_0 = \beta\left(\frac{\partial^2
E}{\partial Q^2}\right)_0 = \beta \left(\frac{\partial \phi}{\partial
Q}\right)_{S,A} = \frac1T\frac{1}{\chi_{S,A}}$, where
Eq.~\eqref{eq:extremaas1law1} was used and the electric susceptibility
$\chi_{S,A}$ has been defined as $\chi_{S,A} = \left(\frac{\partial
Q}{\partial \phi}\right)_{S,A}$.
In summary, the connection of the second derivatives of the action
evaluated at the extrema with thermodynamics coefficients, i.e.,
laboratory variables, is
\begin{align}
&\left(\frac{\partial^2 I_*}{\partial S^2}
\right)_0 = \frac{1}{C_{A,Q}}\,,
\label{eq:derintermsofthermoSS2}
\\
&\left(\frac{\partial^2 I_*}{\partial Q^2}
\right)_0 =
\frac{\beta}{\chi_{S,A}}\,,
\label{eq:derintermsofthermoQQ2}\\
&
\left(\frac{\partial^2
I_*}{\partial S\partial Q} \right)_0
= -\frac{ \lambda_{T,A}}{C_{A,Q}}\,.
\label{eq:derintermsofthermoSQ2}
\end{align}
Therefore, with
Eqs.~\eqref{eq:derintermsofthermoSS2}-\eqref{eq:derintermsofthermoSQ2},
the condition of stability, Eq.~\eqref{eq:condstab2app22}, in
thermodynamic coefficients is $\beta C_{A,Q}^{-1} \chi_{S,A}^{-1} -
\lambda_{T,A}^2 C_{A,Q}^{-2}>0 $. Considering that $C_{A,\phi}^{-1} =
C_{A,Q}^{-1} - T \lambda_{T,A}^2 C_{A,Q}^{-2}$, where the heat
capacity $C_{A,\phi}$ at constant area and electric potential is
defined as $C_{A,\phi} = \left(\frac{\partial S}{\partial
T}\right)_{A,\phi}$, the stability condition becomes $\beta
C_{A,\phi}^{-1} \chi_{S,A}^{-1} >0$. But, because $\chi_{S,A}>0$ in the
case of this ensemble, then we have that the condition reduces to
\begin{align}
C_{A,\phi} >0\,,
\end{align}
with
\begin{align}
&C_{A,\phi} = \frac{(d-2) \Omega r_+^{d-2}} 
{2\left((d-1)x^{d-3}-(d-3)x^{3-d}y-2\right)}\,,
\label{heatcapacityPhiA}
\end{align}
Since one has to impose
$x^{2d-6}>y$ in order to have black holes,
and since $y=\gamma x^{2d-4}$, we recover the condition
Eq.~\eqref{eq:stabilityx}. Moreover, by using that 
$\gamma = \frac{\Phi^2}{x^2 - (1-\Phi^2)x^{d-1}}$
from Eq.~\eqref{eq:extremaaction},
we recover the expression of the heat capacity given by 
Eq.~\eqref{eq:heatcapacityPhibetter}
and the thermodynamic stability condition 
Eq.~\eqref{eq:cphia}.

\section{The case $d=4$ in the zero loop approximation
analyzed by Braden, Brown, Whiting, and York \cite{Braden:1990}}
\label{app:d4case} 

\subsection{Electric charge and potential}

The grand canonical ensemble of a Reissner-Nordstr\"om black hole
in a cavity in $d=4$ was constructed and analyzed in
\cite{Braden:1990}. In this Appendix, we make the comparison of
our results with \cite{Braden:1990} by setting $d=4$ in our
expressions. In order to do this, one must keep in mind that we use
different definitions for the Lagrangian of the electromagnetic
part. In $d=4$, the charge we consider is $q = \sqrt{4\pi} \,q_{\rm B}$ and
the potential we consider is $A_\tau = \frac{A_{\tau\rm
B}}{\sqrt{4\pi}}$, where $q_{\rm B}$ and $A_{\tau\rm B}$ are the electric
charge and the electric potential used in \cite{Braden:1990},
respectively. With
these redefinitions, our analysis for the particular dimension $d=4$
yields the same results as \cite{Braden:1990}.

\subsection{Action and stationary points}

The action Eq.~\eqref{eq:actionI0full} 
for $d=4$ turns into
\begin{align}
I = \beta R \left(1 - f(\beta, \phi, R) \right)
- q \beta \phi - \pi
r_+^2\,,
\label{action4d}
\end{align}
with
\begin{align}
f = 1 - \frac{r_+}{R} - \frac{q^2}{4\pi r_+ R} + \frac{q^2}{4
\pi R^2}\,.
\label{f4d}
\end{align}
From the action gin in Eq.~\eqref{action4d}, one can obtain
the extrema conditions as
given in
Eqs.~\eqref{eq:extremaaction} and \eqref{eq:extrema2action}
for $d=4$. 
Putting $x= \frac{r_+}{R}$, $y = \frac{q^2}{4\pi
R^2}$,  $\gamma =\frac{(4\pi RT)^2\Phi^2}{(1-\Phi^2)^2}$
which is Eq.~\eqref{eq:definitions2} for $d=4$,
Eqs.~\eqref{eq:extremaaction} and \eqref{eq:extrema2action}
yield
\begin{align}
&(1 - \Phi^2)\left(\frac{r_+}{R}\right)^3 -
\left(\frac{r_+}{R}\right)^2 + \frac{(1 - \Phi^2)^2}{(4\pi RT)^2}
= 0\,,
\label{eq:extremaaction4d}
\\
&
\frac{q^2}{4\pi
R^2}=\frac{(4\pi RT)^2\,\Phi^2}{(1 - \Phi^2)^2}
\left(\frac{r_+}{R}\right)^4 \,.
\label{eq:extrema2action4d1}
\end{align}
Now, Eqs.~\eqref{eq:extremaaction4d} and 
\eqref{eq:extrema2action4d1} give precisely 
 Eqs.~(4.15) and (4.16) in \cite{Braden:1990} with the
redefinitions $q = \sqrt{4\pi} q_{\rm B}$ and $\Phi = \phi_{\rm
B}$, where again $q_{\rm B}$ and $\phi_{\rm B}$ are the
electric charge and the electric potential used in
\cite{Braden:1990}.
The equation
for $\frac{r_+}{R}$, Eq.~\eqref{eq:extremaaction4d},
is then a cubic equation, which was solved in
\cite{Braden:1990}.  There are two solutions, $\frac{r_{+1}}{R}$
for the small black
hole and $\frac{r_{+2}}{R}$ for the large black hole.
For each solution,
the electric charge $q$ follows from Eq.~\eqref{eq:extremaaction4d}.

\subsection{Stability condition}

The stability condition in this system for $d=4$ corresponds to
the positive definiteness of the Hessian of the action, meaning that
if a solution of the extrema has positive definite Hessian of the
action, then the solution is stable. The expression of the condition
given in Eq.~\eqref{eq:stabilitysimple}
in $d=4$ is
\begin{align}
\frac{\Bigl(3(1- \Phi^2)\left(\frac{r_+}{R}\right)
-2\Bigr) (1 - \left(\frac{r_+}{R}\right))}
{1 - (1-\Phi^2)\left(\frac{r_+}{R}\right)}
> 0\,.
\label{stabd=4}
\end{align}
The stability condition in Eq.~\eqref{stabd=4} implies then that a 
solution is stable if 
\begin{align}
\frac{r_+}{R}>\frac{2}{3 (1-\Phi^2)} \,.
\label{stabd=4conc}
\end{align} 
These stability equations result should be compared with the 
stability condition given in~\cite{Braden:1990}.
Recovering  $x=\frac{r_+}{R}$, note
that 
$\frac{\left(3(1- \Phi^2)x
-2\right) (1 - x)}
{1 - (1-\Phi^2)x}$ of Eq.~\eqref{stabd=4}
reduces to 
$\frac{
(3x-2)(1 - (1-\Phi^2)x)-\Phi^2x}
{1 - (1-\Phi^2)x}=
3x-2-
\frac{\Phi^2x}
{1 - (1-\Phi^2)x}
$.
Now, from Eq.~\eqref{eq:extremaaction4d}
one has $\frac{1}{1 - (1-\Phi^2)x}=
\frac{(4\pi RT)^2x^2}{(1 - \Phi^2)^2} 
$, so that making this substitution on
the previous equation and multiplying it by $x$
one gets 
$3x^2-2x-
\frac{(4\pi RT)^2\,\Phi^2}{(1 - \Phi^2)^2}
x^4=3x^2-2x-y$, where 
Eq.~\eqref{eq:extrema2action4d1} was used
to recover $y$. 
Thus, Eq.~\eqref{stabd=4} is
the same as 
$3x^2-2x-y\geq 0$.
This latter equation is the stability condition
Eq.~(5.18) in~\cite{Braden:1990}.
Its solution is $x\geq\frac13+\frac13\sqrt{1+3y}$, i.e.,
$\frac{r_+}{R}\geq\frac13+\frac13\sqrt{1+3 \frac{q_{\rm B}^2}{R^2}}$.
This equation and Eq.~\eqref{stabd=4conc} reduce
to $\frac{r_+}{R}\geq\frac23$
when $q_{\rm B}=0$ or $\Phi=0$, i.e., to 
York's case \cite{York:1986}.
Therefore, 
our results agree with~\cite{Braden:1990}. From the two solutions
that exist in the ensemble,
$\frac{r_{+1}}{R}$ is unstable, and 
indeed only the larger black hole solution $\frac{r_{+2}}{R}$ is stable.

\subsection{Thermodynamics}

The thermodynamic quantities obtained for $d=4$ from
Eqs.~\eqref{eq:entropyrp}-\eqref{eq:Energy} are the same as the
thermodynamic quantities obtained in \cite{Braden:1990}, with the
redefinitions $q = \sqrt{4\pi} q_{\rm B}$ and $\Phi =
\phi_{\rm B}$.
To be complete, we list here
those thermodynamic quantities 
in $d=4$, such as the entropy, electric
charge, surface pressure, thermodynamic energy, and area.
One can also find the Euler equation, the Gibbs-Duhem relation,
and the Smarr formula for this case.

The entropy can be directly obtained from Eq.~\eqref{eq:entropyrp},
giving
\begin{align}
S = \frac14 A_+\,,
\label{eq:entropyrp4d}
\end{align}
which is the Bekenstein-Hawking entropy of a four-dimensional black
hole, which explicitly gives $S = \pi r_+^2$, since $A_+=4\pi
r_+^2$.

The electric charge can be computed from Eq.~\eqref{eq:meanQ}, for
$d=4$. The charge has the same appearance as in general $d$, i.e.,
for given
$T$, $\phi$, and $R$, one has $Q=q$, which is the
electric charge of
Reissner-Nordstr\"om space.

The thermodynamic surface pressure at $R$ can be
calculated from Eq.~\eqref{eq:meanpressure} for the $d=4$ case, giving
\begin{align}
 p = \frac{1}{16\pi R \sqrt{f}}&\left(\left(1-
\sqrt{
\left(1-\frac{r_+}{R}\right) \left(1-\frac{
q^2}{4\pi r_+ R}\right)}
\right)^2
\right.\nonumber
\\
&\Big.
-\frac{q^2}{4\pi R^{2}}\Biggr)\,,
\end{align}
where  $f$ is given
in Eq.~\eqref{f4d}.
This tangential surface pressure acts along an
area $A=4\pi R^2$ of the boundary of the cavity.

The mean thermodynamic energy
is given in Eq.~\eqref{eq:Energy}, which in $d=4$ case is
\begin{align}
E = R
\left(1-
\sqrt{
\left(1-\frac{r_+}{R}\right) \left(1-\frac{
q^2}{4\pi r_+ R}\right)}
\right)\,.
\label{eq:Energy4d}
\end{align}
This is the same expression as the quasilocal energy evaluated at a
spherical shell of radius $R$.

The first law of thermodynamics,
$TdS=dE+pdA-\phi dQ$ given in Eq.~\eqref{1stlawusual},
for the system in $d=4$ holds.

Moreover, from Eq.~\eqref{eq:Energy4d}, 
one can write the energy
in terms of the
entropy $S$ of Eq.~\eqref{eq:entropyrp4d},
electric charge $Q$, and
surface area of the cavity $A$, 
as
\begin{align}
E =&
\left(\frac{A}{4\pi}\right)^{\frac{1}{2}}\times
\nonumber\\
&\left(1-\sqrt{\left(1-\left(\frac{4
S}{A}\right)^{\frac{1}{2}}\right)\left(1-
\frac{Q^2}{(4SA)^{\frac{1}{2}}}\right)}
\right)\,.
\label{eq:energyintermsof4d}
\end{align}
One can then use the Euler's homogeneous function
theorem considering that under rescaling of its arguments, the energy
as a function has the property that $E\left(\nu S,\nu A,\nu
Q^{2}\right) = \nu^{\frac{1}{2}} E\left(S,A,
Q^{2}\right)$. We thus have an integrated version of the
first law of thermodynamics given by, see
Eq.~\eqref{eq:1stlawintegrated},
\begin{align}
E = 2(TS - pA) + \phi
Q\,,\label{eq:1stlawintegrated4d}
\end{align}
which is the Euler equation for
a $d=4$
electrically charged black hole in a
heat reservoir.
By differentiating Eq.~\eqref{eq:1stlawintegrated4d} and considering
that $dE = TdS - pdA + \phi dQ$, one obtains
\begin{align}
  TdS - pdA + 2(SdT-Adp) + Qd\phi = 0\,.
\label{eq:gibbs4d}
\end{align}
which is the Gibbs\hskip0.04cm-Duhem relation for
this case.
Then, the Smarr formula in $d=4$ is 
\begin{align}
m = 2 T_{\rm H}S + \phi_{\rm H}Q\,,
\label{eq:smarr4d}
\end{align}
see Eq.~\eqref{eq:smarr}. Again, the Smarr
formula is valid for the small black hole
solution only.

The stability condition is related to thermodynamics as it
corresponds to the positivity of the heat capacity at constant area
and electric potential, which we show for arbitrary dimension and it
was shown in \cite{Braden:1990} for $d=4$.

\subsection{Most probable or most favorable configurations}

Again, the most probable, or most favorable,
configurations
are  found from 
the state with the lowest value of the
action $I_0$,
or what amounts to the same thing,
the state with the lowest value of the
grand potential $W$.
In $d=4$ this comparison was not performed in
\cite{Braden:1990}.

First, we have to compare the large stable black hole $r_ {+2}$ with
hot flat space given by the stationary point $r_+ = 0$ and $q = 0$,
which in $d=4$ and contrarily to $d>4$, is not smooth.  To simulate
hot flat space, a hot conductor charged sphere is considered, which from
Eq.~\eqref{Ihotflatsphere} it has an action
$I_{\rm hot \,flat\,sphere} = - \frac{2\pi} { \frac{1}{r_{\rm hs}}-
\frac{1}{R}} \beta\phi^2$ for $d=4$.  For tiny spheres, which are the ones that
better simulate hot flat space, the action for the stable black hole
Eq.~\eqref{action4d} is less than the action for $I_{\rm hot
\,flat\,sphere}$, and so the stable black hole dominates.

Second, we have to compare the large stable black hole $r_ {+2}$ 
with the other 
 stationary point, which in $d=4$ is  $r_+=R$ and
$\frac{q}{\sqrt{4\pi}}=R$. One finds
that the stable black hole is a more probable
configuration than the extreme black hole with
the horizon at the cavity.

\section{Null geodesic sphere of a $d$-dimensional Reissner-Nordstr\"om
spacetime}
\label{photonsphere}

The $d$-dimensional Reissner-Nordstr\"om spacetime has
the line element given
by
\begin{align}
ds^2 = - f(r)\,dt^2 + \dfrac{dr^2}{f(r)} + r^2 d\Omega^2\,,
\end{align}
where
\begin{align}
f(r) = 1 - \frac{2\mu m}{r^{d-3}} + \frac{\lambda q^2}{r^{2d-6}}\,,
\label{fofr}
\end{align}
with $\mu = \frac{8\pi}{(d-2)\Omega}$, and 
$d\Omega$ is the surface element of a $d-2$-sphere
with unit radius given by
\begin{align}
d\Omega^2 = d\varphi_0^2 +  
\sum_{i=1}^{d-3}\left(\prod_{j=0}^{i-1}\sin^2\varphi_j\right)
d\varphi_i^2\,.
\end{align}

We now consider a congruence of null geodesics with tangent vector
$l^a = \frac{d x^a}{d \nu}$, where $\nu$ is the parameter of the
geodesic. Since the Reissner-Nordstr\"om spacetime
is static and
possesses spherical
symmetry, one can integrate the equations of motion
using integration constants. One integration constant
is related to the spacetime being
static, which is characterized by the Killing vector $\xi =
\partial_t$, and gives the notion of
the specific energy of the geodesic $\varepsilon$, yielding
\begin{align}
\Dot{t}=\frac{\varepsilon}{f(r)}\,.
\label{econst}
\end{align}
Another integration constant 
is the specific angular
momentum  which is conserved. Due to spherical
symmetry, it is possible to consider only the equatorial geodesics
without loss of generality, i.e., we can fix $\varphi_i = \frac{\pi}{2}$
for $i=0, ..., d-4$. Therefore, the specific angular momentum $h$
is given
by
\begin{align}
\Dot{\varphi}_{d-3}=\frac{h}{r^2}\,.
\label{hconst}
\end{align}
Finally, the remaining integration constant is the fact that the
vector tangent to the geodesic is null, i.e.,
$
g_{ab}l^a l^b = 0$.
This condition becomes
\begin{align}
\Dot{r}^2 = f(r) \Dot{t}^2 -
r^2\Dot{\varphi}_{d-3}^2
\,.
\label{0const}
\end{align}
Using, Eqs.~\eqref{econst} and
\eqref{hconst} in \eqref{0const}
we have
\begin{align}
\Dot{r}^2 = \varepsilon^2 - \frac{h^2}{r^2}
\left(1 - \frac{2\mu m}{r^{d-3}} + \frac{\lambda Q^2}{r^{2d-6}}
\right)
\,,
\label{rdotsquared}
\end{align}
where Eq.~\eqref{fofr} has also been used.
Note now, that from Eq.~\eqref{hconst}
one has
$\frac{\Dot{r}}{h}=-\frac{d}{d\phi}(\frac{1}{r})$.
So Eq.~\eqref{rdotsquared} turns into
\begin{align}
\left(\frac1r\right)'^2 = \frac{\varepsilon^2}{h^2} -
\left(\frac1r\right)^2
+ 2\mu m \left(\frac1r\right)^{d-1} - \lambda q^2
\left(\frac1r\right)^{2d - 4}\,,
\label{rphisquared}
\end{align}
where here a prime denotes differentiation with respect to
$\phi$.
Differentiating once with respect to
$\phi$, we have the equation
\begin{align}
\left(\frac1r\right)^{\prime\prime} + \frac1r =
\mu m (d-1) \left(\frac1r\right)^{d-2} - \lambda Q^2 (d-2)
\left(\frac1r\right)^{2d-5}
\label{rphiphi}
\end{align}
for null geodesics.

The null geodesic sphere is characterized by circular null geodesics,
therefore $r$ is constant, i.e., $r'=0$
and $r''=0$. From the geodesic equation,
Eq.~\eqref{rphiphi}, the value of
$r$ of the radius of a circular
null geodesic must then obey the equation
\begin{align}
1 = \mu m (d-1) \left(\frac1r\right)^{d-3} - \lambda q^2 (d-2)
\left(\frac1r\right)^{2(d-3)}\,.
\label{geodphotoncircular}
\end{align}
The solutions for this radius, for each dimension $d$,
are thus functions of $m$ and $q$, i.e., $r=r(m,q)$.
But as we have seen $m=m(r_+,q)$, explicitly, 
\begin{align}
2\mu m = r_+^{d-3} + \frac{\lambda q^2}{r_+^{d-3}}\,,
\label{mr+qd}
\end{align}
which is found from the zero of Eq.~\eqref{fofr}.
Moreover, from 
Eq.~\eqref{eq:phiexpress} and 
Eq.~\eqref{eq:definitions},
we can define formally an electric potential
$\Phi$ at the null geodesic circular
radius $r$ by the expression
\begin{align}
\Phi(r_+,q,r)=
\frac{ \sqrt{\lambda}q}{
\sqrt{f[R,r_+,q]}}\left(\frac{1}{r_+^{d-3}}-
\frac{1}{r^{d-3}}\right)
\label{Phigenerized}
\end{align}
which can be inverted to
$q=q(r_+,\Phi,r)$. So
also $m=m(r_+,\Phi,r)$.
Indeed, from Eqs.~\eqref{mr+qd}
and \eqref{Phigenerized} we get the correspondence
\begin{align}
&\mu m = \frac{1}{2}\left( 1 +
\frac{\Phi^2}
{1 -
(1 - \Phi^2)
\left(\frac{r_+}{r}\right)^{d-3}}\right)\,r_+^{d-3}
\,,
\label{mum1}
\\
&\lambda q^2 =\quad\;
\frac{\Phi^2}{1 - (1 - \Phi^2)\left(\frac{r_+}{r}\right)^{d-3}
}\;\; r_+^{2(d-3)}
\,.
\label{lambdaq1}
\end{align}
Using Eqs.~\eqref{mum1} and \eqref{lambdaq1}
in the geodesic equation, Eq.~\eqref{geodphotoncircular},
one has
\begin{align}
&1 - \left(\frac{d+1}{2} + \frac{d-3}{2}\Phi^2\right)
\left(\frac{r_+}{r}\right)^{d-3}
\nonumber\\&+ \left(\frac{d-1}{2} + \frac{d-3}{2}\Phi^2
\right)\left(\frac{r_+}{r}\right)^{2(d-3)} = 0,
\end{align}
which has the solution $r=r_{\rm ps}$ with
\begin{align}
r_{\rm ps}=\left[
\frac
{(d-1)\left(1 +
\frac{d-3}{d-1}\Phi^2\right)}{2}\right]^\frac{1}{d-3}
r_+ \,.
\end{align}
This radius $r_{\rm ps}$ is the radius of a circular
null geodesic in a Reissner-Nordstr\"om black hole
geometry with given $r_+$, $\Phi$ and $d$.
If photons, gravitons, or any
other lightlike particle, are
placed at this null geodesic they will follow
circular orbits. The corresponding
sphere radius is usually called
photon sphere radius, $r_{\rm ps}$.


\end{document}